\documentclass{article}

\usepackage[a4paper, total={6in, 10in}]{geometry}
\usepackage{authblk}

\usepackage{graphicx} 
\usepackage{hyperref}       
\usepackage{color}
\usepackage[numbers, compress]{natbib}
\usepackage{amsmath}
\usepackage{amsfonts}
\usepackage{amssymb}
\usepackage{comment}
\usepackage{xcolor}
\usepackage{algorithm}
\usepackage{algpseudocode}
\usepackage{subcaption}
\usepackage{caption}
\usepackage{graphicx}
\usepackage{booktabs}
\usepackage{pdfpages}

\newcommand\Flip{\mathrm{Flip}}

\newcommand\uscore[2]{\mathbf{#1}_{\text{#2}}}

\DeclareMathOperator*{\argmax}{argmax}

\title{Benchmarking Graph Neural Networks in Solving Hard Constraint Satisfaction Problems}
\author[1,2]{G. Skenderi}
\author[3]{L. Buffoni}
\author[4,5]{F. D'Amico}
\author[4,5,6]{D. Machado}
\author[3]{R. Marino}
\author[4,5]{M. Negri}
\author[4,5,7]{F. Ricci-Tersenghi}
\author[1,2]{C. Lucibello}
\author[4,7]{M. C. Angelini} 

\affil[1]{Department of Computing Sciences, Bocconi University, Milan, Italy}
\affil[2]{Bocconi Institute for Data Science and Analytics (BIDSA)}
\affil[3]{Department of Physics and Astronomy and INFN, University of Florence}
\affil[4]{Dipartimento di Fisica, Sapienza Università di Roma}
\affil[5]{CNR - Nanotec, unità di Roma, P.le Aldo Moro 5, 00185 Rome, Italy}
\affil[6]{Department of Theoretical Physics, Physics Faculty, University of Havana. CP10400, La Habana, Cuba}
\affil[7]{INFN, sezione di Roma1, P.le Aldo Moro 5, 00185 Rome, Italy}

\begin{document}

\maketitle 

\begin{abstract}
Graph neural networks (GNNs) are increasingly applied to hard optimization problems, often claiming superiority over classical heuristics. However, such claims risk being unsolid due to a lack of standard benchmarks on truly hard instances. From a statistical physics perspective, we propose new hard benchmarks based on random problems. We provide these benchmarks, along with performance results from both classical heuristics and GNNs. Our fair comparison shows that classical algorithms still outperform GNNs. We discuss the challenges for neural networks in this domain. Future claims of superiority can be made more robust using our benchmarks, available at \url{https://github.com/ArtLabBocconi/RandCSPBench}.
\end{abstract}


\section{Introduction}

Constraint satisfaction problems (CSPs), a specific subset of optimization problems, play a central role in many fields of scientific research and technological applications. A CSP is defined by a set of variables and a set of constraints, the goal being to determine whether there exists a \emph{satisfying assignment} for the variables that solves all the constraints (\textit{decision problem}) and eventually to find such an assignment (\textit{assignment problem}).

In recent years, machine learning (ML) methdos, and in particular graph neural networks (GNNs) \cite{scarselli2008graph}, have been widely applied both as end-to-end solvers \cite{holden_machine_2021,shi_satformer_2022,selsam_learning_2019,li_deepsat_2022,ozolins_goal-aware_2022,yan_addressing_2023,sun_difusco_2023,freivalds_denoising_2022,duan_augment_2022,achlioptas_hide_2022,Karalias2020,Joshi2020,cameron_predicting_2020,Gasse2019,Li2018,li_nsnet_2022, cappart2023combinatorial, kotary2021end, bengio2021machine} or as guides for heuristic moves in traditional solvers \cite{ciarella2023machine,wang2024neuroback}. 
Despite this surge of activity, there is still no well-defined, unified benchmark for comparing ML-based approaches with classical heuristic algorithms. As a result, many newly proposed methods are evaluated on only a small number of instances drawn from different datasets. This lack of standardization weakens the scientific rigor of the evaluations, making many performance claims difficult to interpret or compare \cite{schuetz2022combinatorial, boettcher2023inability, angelini2023modern, schuetz2023reply, schuetz2023replym}.

However, the problem of identifying classes of hard problems was already faced in the 90s to assess performances of algorithms for optimization problems \cite{cheeseman1991really}. By analyzing ensembles of random instances of a given problem \cite{kirkpatrick1994critical, selman1996generating}, researchers in statistical physics were able to discover and characterize various phase transitions that alter the geometry of the solution space. These transitions provide insight into how hard an instance of a problem is likely to be. They appear at well-defined values of some parameters, and influence the behavior of classical heuristic algorithms \cite{monasson1999determining, krzakala2007gibbs,ricci2009cavity, mezard2009information, marino2023hard, angelini2023limits}.
In particular, the emergence of a glassy energy landscape has been shown to hinder the performance of Monte Carlo algorithms. Similarly, the appearance of frozen solutions poses significant obstacles for even the most advanced message-passing algorithms \cite{zdeborová2009statistical}. At the same time, simpler heuristics, such as greedy methods or gradient-descent-like algorithms, may fail even before the phase transitions take place.

In light of these findings, a natural question arises: \textbf{Are neural solvers subject to the same structural barriers that affect classical heuristic algorithms, or do they exhibit fundamentally different failure modes?}

Recently, it has been shown that classical heuristic algorithms can work in different regimes depending on how the running time is scaled with the problem size \cite{angelini2025algorithmic}. \textbf{How does the performance of a neural solver scale with its size?}

In this paper, we address the above questions by introducing new benchmarks inspired by the statistical physics framework.

A recent work \cite{li_g4satbench_2023} introduced a dataset aimed at comparing ML approaches to CSPs. The dataset is organized into three difficulty levels, defined primarily by problem size, and the proposed evaluation framework is specifically tailored to GNN-based solvers for satisfiability problems. By contrast, our dataset not only includes problems of varying sizes, but also exploit a notion of incremental hardness, linked to problem-specific parameters, such as the connectivity of the underlying graphs. As a result, our benchmark provides a richer and more diverse set of instances, systematically organized to enable a more fine-grained and meaningful evaluation of algorithmic performance as a function of problem difficulty. Our new dataset focuses on two well-known CSPs, the $K$-SAT problem and the $q$-coloring ($q$-col) problem and we compare different classical algorithms with GNN-based SAT solvers. 

The $K$-SAT problem asks whether there exists an assignment of $N$ Boolean variables that satisfies a conjunction of $M$ clauses (constraints), where each clause is a disjunction of exactly $K$ literals, and each literal corresponds to a variable or its negation. In the random $K$-SAT problem, every clause contains $K$ randomly chosen literals, corresponding to different variables. The random $q$-col problem asks for an assignment of colors to the $N$ nodes of an undirected Erd\H{o}s-Renyi random graph with mean degree $c$ (thus, the number of constraints is $M=cN/2$). Each node must be assigned one of $q$ possible colors, under the constraint that no two neighboring nodes have the same color.

Varying control parameters $\alpha = M/N$ in $K$-SAT and $c$ in $q$-col, it has been shown that the structure of the solution space undergoes a series of phase transitions. As $\alpha$ or $c$ increases, the uniform measure over the space of satisfying assignments breaks up into an exponential number of pure states at the \textit{clustering} transition $\alpha_d(K)$ or $c_d(q)$, and subsequently condenses onto a subexponential number of dominant states at the \textit{condensation} transition $\alpha_c(K)$ or $c_c(q)$ \cite{krzakala2007gibbs}.
Finally, there is a sharp transition between satisfiable and unsatisfiable phases at the \textit{satisfiability threshold} $\alpha_s(K)$ or $c_S(q)$. All thresholds depend only on $K$ or $q$ in the limit $N \to \infty$. These phase transitions have been rigorously established in \cite{Ding2022annalen}.
By tuning such parameters, we can generate ensembles of instances with varying levels of difficulty and assess the algorithmic performance across different regimes of problem hardness. 

For $K=3$ in the $K$-SAT problem, $\alpha_d=\alpha_c$ and the split of the space of solution is of a different type \cite{Montanari_2008}, and the same happens for the $q$-col problem for $q=3$. In the statistical mechanics language, the case $K\ge4$ ($q\ge 4$) is described by the 1-step replica symmetry breaking (1RSB) theory while for $K=3$ ($q=3$) 1RSB theory is not enough and one should probably use a full-RSB theory \cite{mezard1987spin, mezard2009information}. Both problems, with $K=3$ and $K\ge 4$ are NP-hard; however, while classical algorithms can find a satisfiable assignment for a typical instance of the random problem quite near to $\alpha_s$ in the $K=3$ case \cite{marino_backtracking_2016}, for $K\ge4$, approaching closely $\alpha_s$ seems impossible with existing algorithms. Unfortunately, previous works applying ML tools to $K$-SAT ($q$-col) problems mainly focused on the $K=3$ ($q=3$) case, while new algorithms should be tested also on the difficult cases $K>3$ ($q>3$). Our dataset is innovative in this sense, including both 4-SAT and 5-col instances. 

Among classical algorithms we analyze: Monte Carlo algorithms, such as Simulated Annealing (SA); Message Passing (MP) algorithms with decimation, such as Belief Propagation (BP) for $q$-col and Survey Propagation (SP) for $K$-SAT; Local Search algorithms, like Focused Metropolis Search (FMS). Among the available GNN-based solvers in the literature, we focus on NeuroSAT and QuerySAT \cite{selsam_learning_2019,ozolins2022goal} for $K$-SAT and a Physics-Inspired GNN \cite{Schuetz_PIGNNcol_2022} with a recurrent update (rPI-GNN) for $q$-col. Furthermore, we train QuerySAT on the SAT-reduced col problems, in order to make  observations across problems and architectures. 
These architectures represent well-known paradigms for the two specific CSPs we consider in this work, and are among the most well-known approaches in the respective literatures. In particular, NeuroSAT has been featured as the only SAT-specific solver in the recent benchmark paper of Li et al. \cite{li_g4satbench_2023}, and QuerySAT provides a natural extension to this architecture. On the other hand, PI-GNNs have experienced a growing research interest in combinatorial optimization \cite{schuetz2022combinatorial}. Therefore, we use a particular instantiation of these models for the $q$-coloring problem. Last but not least, all the aforementioned architectures have problem-specific designs, but they are based on well-established and common deep learning design principles, which greatly helps our analysis. In this way, we are able to focus mostly on the algorithm class without worrying excessively about the choice and tuning of hyperparameters. The full details on the implementation of all the algorithms are reported in Sec. \ref{sec:methods}

\section{Results}

\subsection{Benchmark datasets}\label{sec:main-datasets}

Our benchmark comprises instances for two canonical classes of CSPs: $K$-SAT and $q$-col. We consider $K \in \{3, 4\}$ and $q \in \{3, 5\}$ to systematically explore increasing levels of complexity. Each instance is designed for the assignment problem. If no solution is found, algorithms are expected to return the configuration that minimizes the total energy (i.e., the number of unsatisfied constraints).

For both $K$-SAT and $q$-col problems, we sample parameters across critical regions. In particular, the clause-to-variable ratio $\alpha$ for SAT (and constraint density $c$ for coloring) is chosen to span several values around the satisfiability threshold. This selection ensures a mixture of satisfiable and unsatisfiable instances of varying hardness. The specific parameter ranges and satisfiability thresholds are:
\begin{itemize}
    \item $\alpha \in [3, 5]$ with $\Delta\alpha = 0.1$ for 3-SAT ($\alpha_s=4.267$),
    \item $\alpha \in [8, 10]$ with $\Delta\alpha = 0.1$ for 4-SAT ($\alpha_s=9.931$),
    \item $c \in [3.32, 4.94]$ with $\Delta c = 0.18$ for 3-coloring ($c_s=4.687$),
    \item $c \in [9.9, 13.5]$ with $\Delta c = 0.4$ for 5-coloring ($c_s=13.669$).
\end{itemize}

The core datasets use variable sizes $N \in \{16, 32, 64, 128, 256\}$. For every $(N, \alpha)$ combination, 1600 training and 400 testing instances are created for 3-SAT, while for 4-SAT we generate only 800 training and 200 testing instances, due to the increased instance sizes.  

To assess the ability of neural solvers to generalize beyond the training regime, we provide an additional test set of out-of-distribution (OOD) instances with larger sizes ($512\le N\le 16384$)
while using the same $\alpha$ values as the in-distribution test dataset. The OOD test dataset is mainly intended as a challenge for the community.

The generation of $q$-col instances follows a similar design philosophy but is performed using a separate pipeline. As with the SAT data, $q$ and $c$ are chosen to explore easy and hard phases of the problem. 
Additional details regarding the dataset generation process are provided in Sec. \ref{sec:methods}.

The total sizes of the training datasets are  
$n_\text{tot}=  168000$ (3-SAT), 
$n_\text{tot}=  84000$ (4-SAT), 
$n_\text{tot}=  80000$ (3-col), 
$n_\text{tot}=  80000$ (5-col). See Sec. \ref{sec:Score} for other statistics of the datasets.

\begin{figure}[]
    \centering
    \begin{subfigure}{0.32\textwidth}
        \centering
        \includegraphics[width=\linewidth]{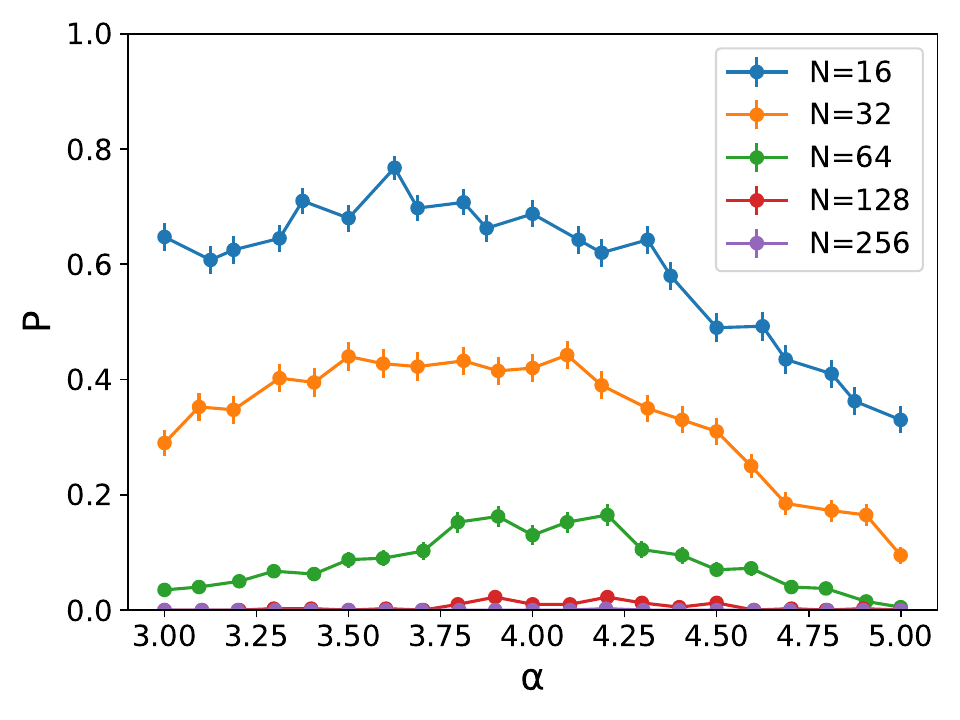}
        \caption{Supervised, 32 iter}
        \label{fig:neurosat_3sat_sup_32}
    \end{subfigure}
    \hfill
    \begin{subfigure}{0.32\textwidth}
        \centering
        \includegraphics[width=\linewidth]{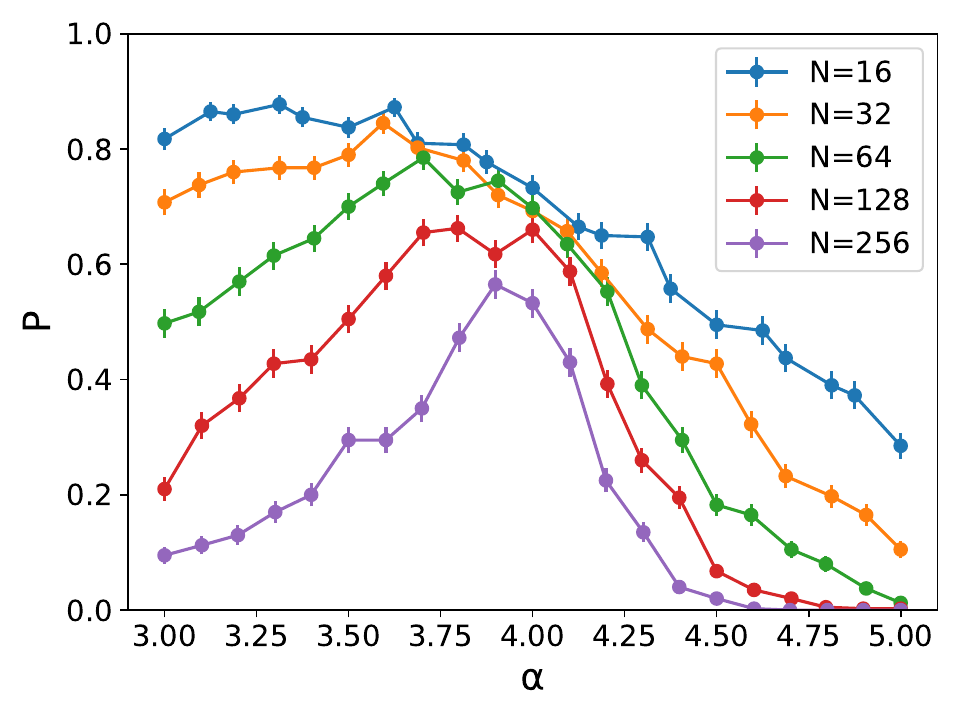}
        \caption{Supervised, 512 iter}
        \label{fig:neurosat_3sat_sup_512}
    \end{subfigure}
    \hfill
    \begin{subfigure}{0.32\textwidth}
        \centering
        \includegraphics[width=\linewidth]{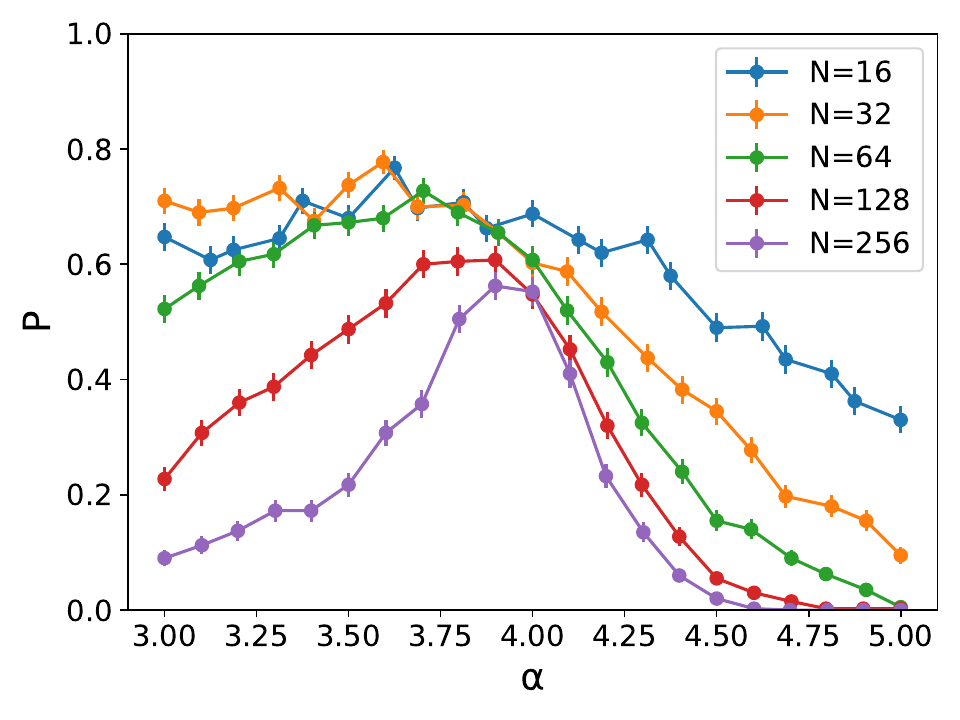}
        \caption{Supervised, $2N$ iter}
        \label{fig:neurosat_3sat_sup_2n}
    \end{subfigure}
    \\
    \vspace{8mm}
    \begin{subfigure}{0.32\textwidth}
        \centering
        \includegraphics[width=\linewidth]{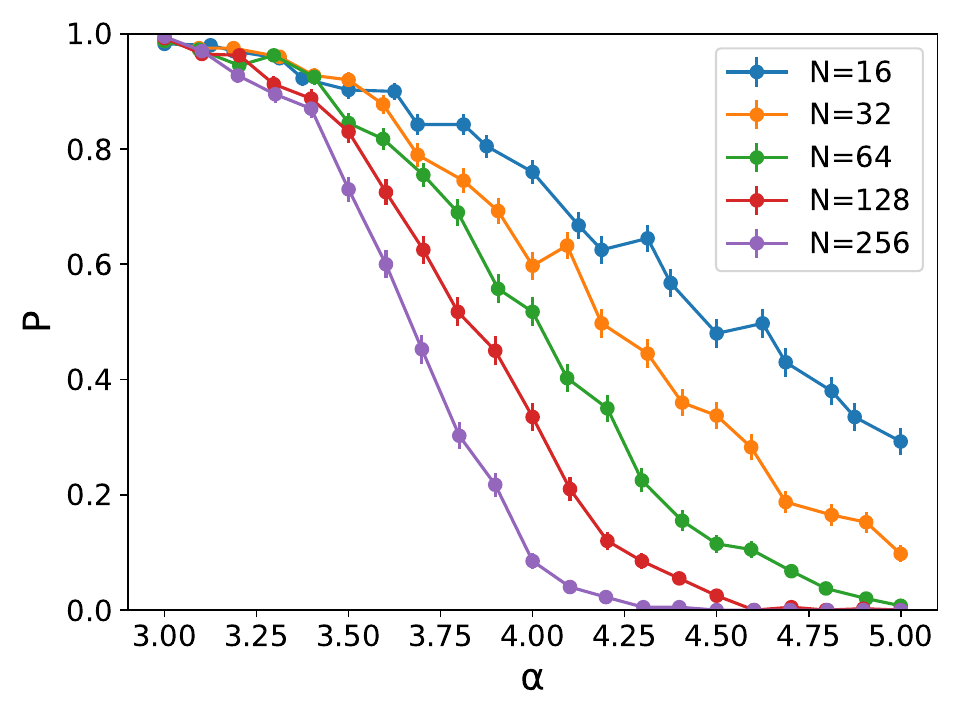}
        \caption{Unsupervised, 32 iter}
        \label{fig:neurosat_3sat_unsup_32}
    \end{subfigure}
    \hfill
    \begin{subfigure}{0.32\textwidth}
        \centering
        \includegraphics[width=\linewidth]{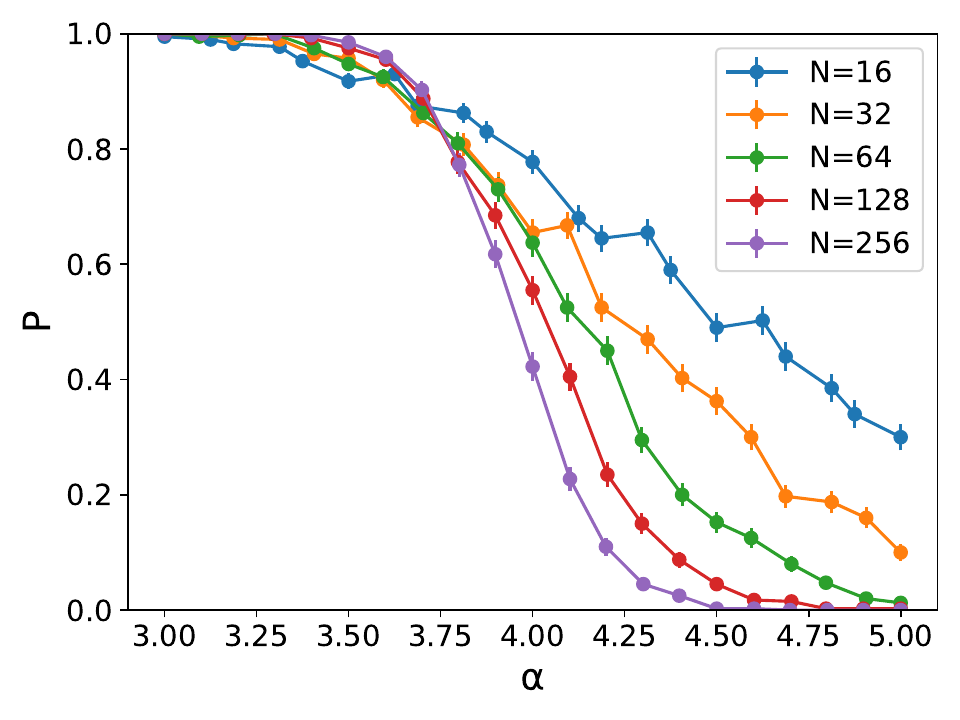}
        \caption{Unsupervised, 512 iter}
        \label{fig:neurosat_3sat_unsup_512}
    \end{subfigure}
    \hfill
    \begin{subfigure}{0.32\textwidth}
        \centering
        \includegraphics[width=\linewidth]{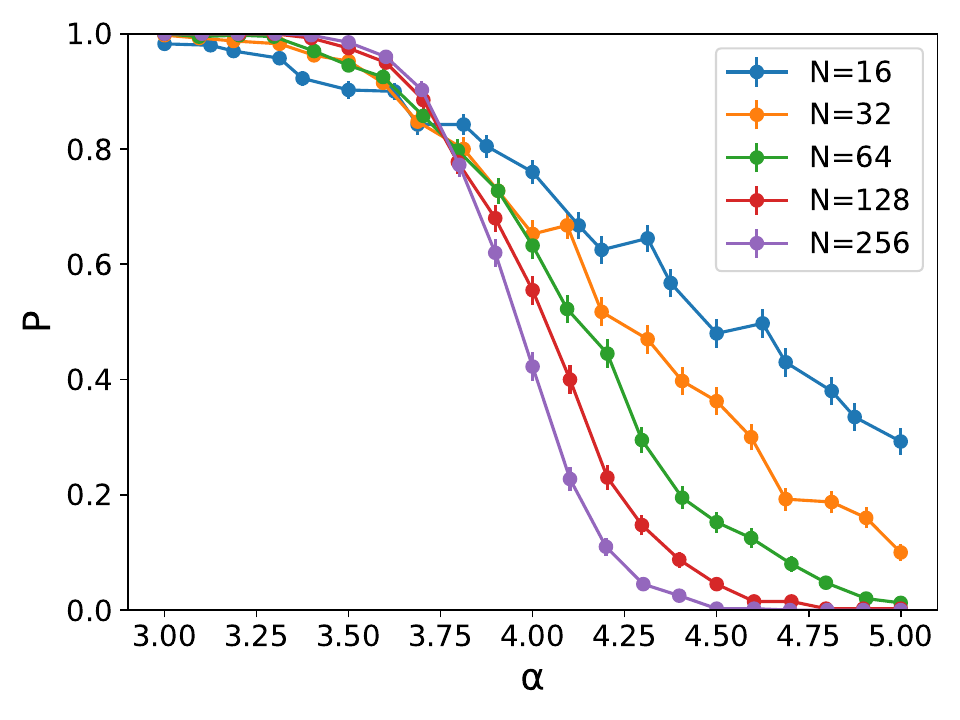}
        \caption{Unsupervised, $2N$ iter}
        \label{fig:neurosat_3sat_unsup_2n}
    \end{subfigure}
    \caption{Probability of finding a satisfying assignment to 3-SAT problems as a function of $\alpha=M/N$ with the supervised (top row) and unsupervised (bottom row) NeuroSAT GNN model. To validate claims regarding the importance of scaling the inference time message-passing iterations of GNNs with the problem size ($N$), we test the model in three different scenarios by performing a small number of fixed iterations (left), a large number of fixed iterations (center), and linearly scaling the number of iterations with the problem size (right). The figure shows that the linear scaling procedure produces excellent results. It is worth noting that with the scaling, inference time is handled more consistently (by definition). For instance, using 512 iterations takes significantly longer for smaller sizes compared to employing linear scaling. Finally, it is clear to see that training an unsupervised model produces much better results compared to the supervised case.}
    \label{fig:neurosat-linscaling-iter}
\end{figure}
 
\subsection{Scaling of times with the size}

We analyze the performances of the algorithms described in sec. \ref{sec:classicAlg} and \ref{sec:GNN} on the benchmark dataset described in Sec.~\ref{sec:main-datasets}. The results crucially depend on the choice of the running time. As noted by the QuerySAT authors \cite{ozolins_goal-aware_2022}, increasing the number of iterations in recurrent neural solvers can lead to significant performance gains. This aligns with recent research on increasing test-time compute to enhance the reasoning capabilities of large language models \cite{snell2024scaling}, including recursion-based strategies \cite{geiping2025scalingtesttimecomputelatent}. 

Building on this idea, we take a step further by scaling the computing times with the problem size.
It is already known that SA and FMS reach larger algorithmic thresholds when they are run on timescales growing with the problem size \cite{angelini2025algorithmic}. For this reason, in this work we run SA (resp.~FMS) with a maximum number of time-steps $t=1000N$ (resp.~$t=100N$) for $K$-SAT and $t=1000N$ (resp.~$t=625N$) for $q$-col. The constant in the linear scaling between $t$ and $N$ is chosen to have the same maximum wall-clock running time for both algorithms, as seen in Sec.~\ref{sec:runningtimes}. Please notice that $t=O(N)$ implies a wall-clock running time growing as $O(N^2)$, as every step takes $O(N)$ operations. Also, message-passing algorithms are run for a number of steps scaling with $N$: indeed, at every decimation step, we fix just one variable, thus requiring $N$ decimation steps. Analogously, we run GNNs for a number of iterations that scales with $N$ only at test-time.

In Fig.~\ref{fig:neurosat-linscaling-iter}, we plot the probability of finding a satisfying assignment for the 3-SAT problem as a function of $\alpha$ for both supervised and unsupervised NeuroSAT GNN model in three different scenarios by performing: (a) a small fixed number (32) of iterations; (b) a large fixed number (512) of iterations; (c) linearly scaling the number of iterations with the problem size as $2N$. The linear scaling returns results equivalent to taking a large number of fixed iterations, as 512 is not smaller than $2N$ for all the sizes studied. However, it will become the best option for large $N$ values (we have verified that this is true also for QuerySAT).

From Fig.~\ref{fig:neurosat-linscaling-iter}, it is also clear that the supervised NeuroSAT solver has very poor performance with respect to the unsupervised one. For this reason, in the following, we will only show results for the unsupervised NeuroSAT, using the linear scaling $t=2N$.

\begin{figure}  
\centering
    \begin{subfigure}{0.48\textwidth}
        \centering
        \includegraphics[width=.8\linewidth]{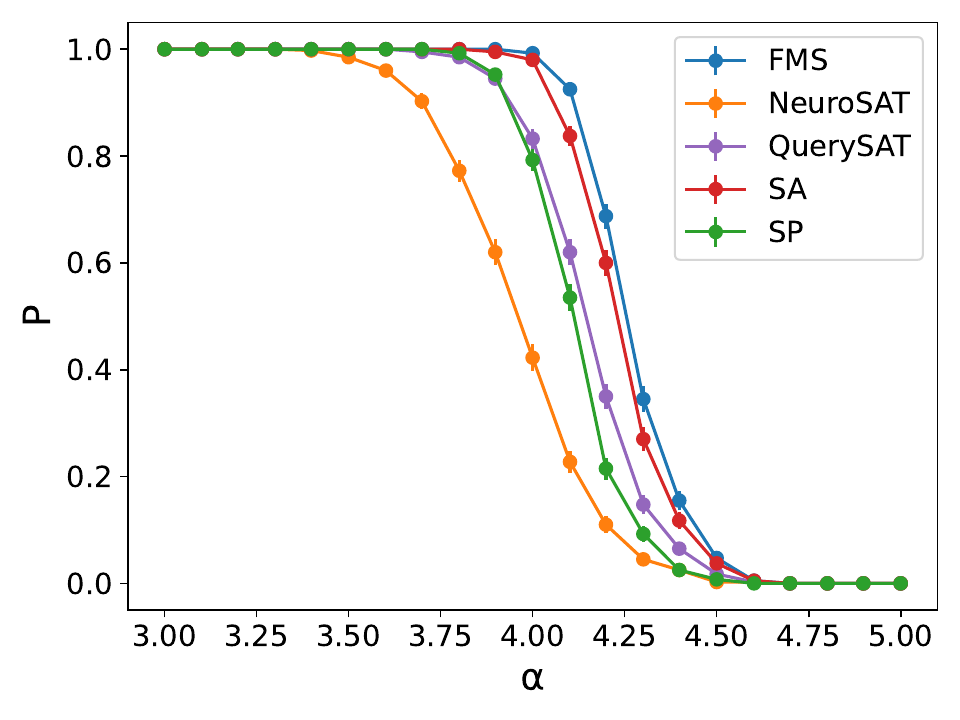}
        \caption{3-SAT $N=256$}
        \label{fig:3sat256}
    \end{subfigure}
    \hfill\vspace{7mm}
    \begin{subfigure}{0.48\textwidth}
        \centering
        \includegraphics[width=.8\linewidth]{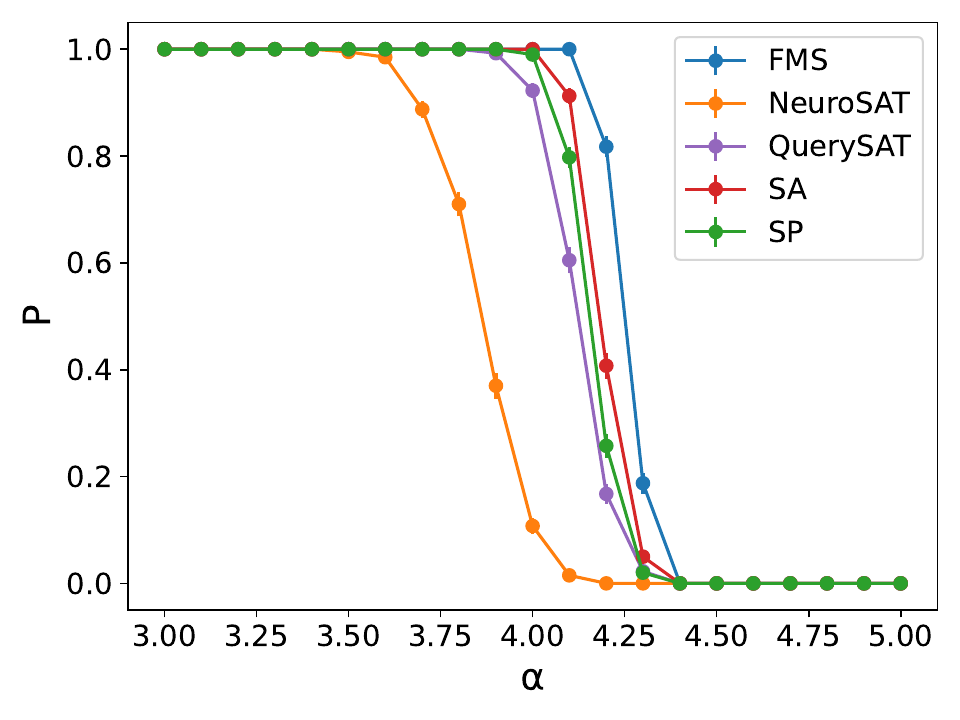}
        \caption{3-SAT $N=1024$}
        \label{fig:3sat1024}
    \end{subfigure}
    
    \begin{subfigure}{0.48\textwidth}
        \centering
        \includegraphics[width=.8\linewidth]{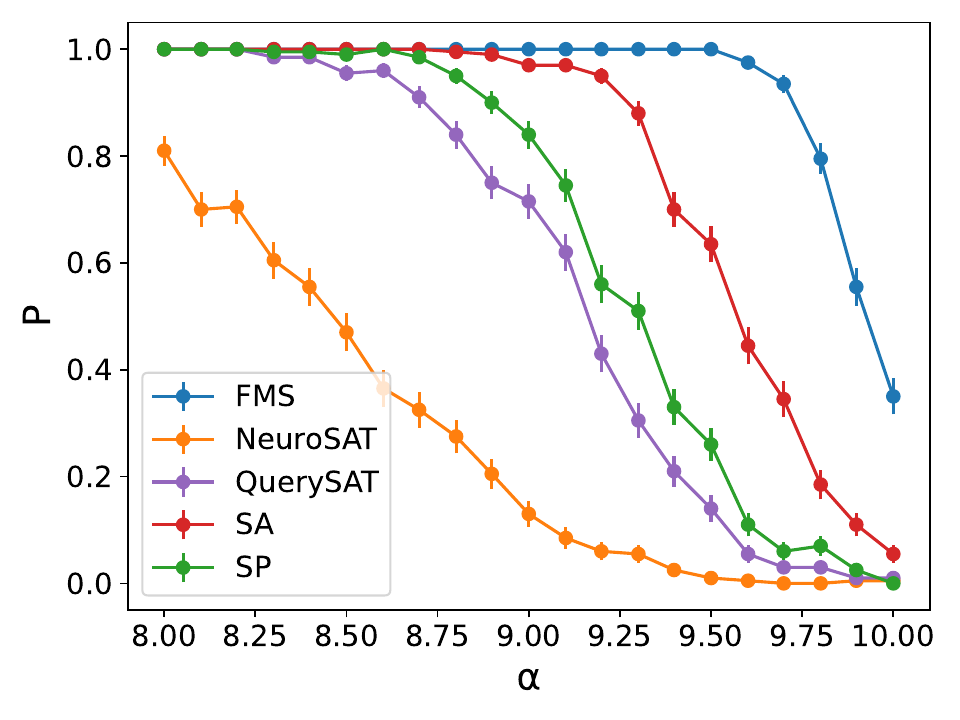}
        \caption{4-SAT $N=256$}
        \label{fig:4sat256}
    \end{subfigure}
    \hfill\vspace{7mm}
    \begin{subfigure}{0.48\textwidth}
        \centering
        \includegraphics[width=.8\linewidth]{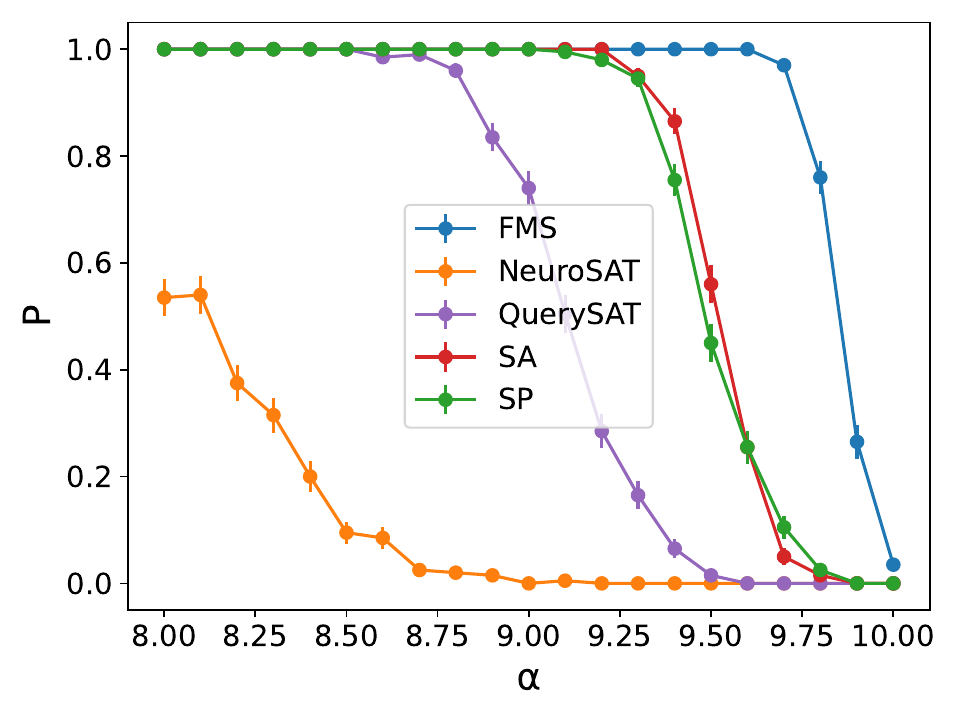}
        \caption{4-SAT $N=1024$}
        \label{fig:4sat1024}
    \end{subfigure}
    
    \begin{subfigure}{0.48\textwidth}
        \centering
        \includegraphics[width=.8\linewidth]{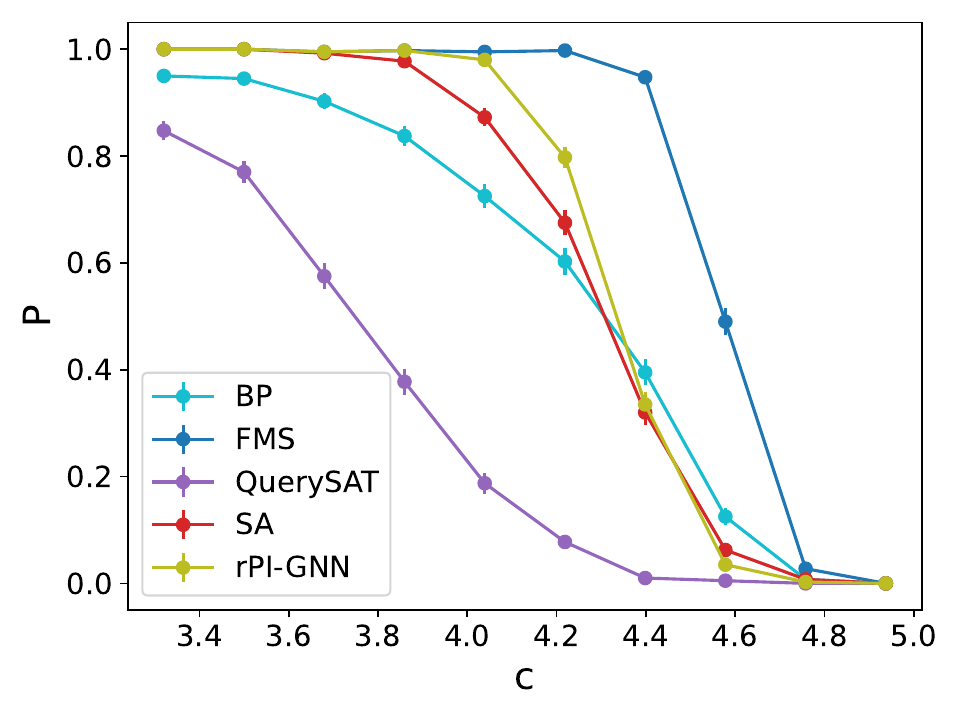}
        \caption{3-col $N=256$}
        \label{fig:3col256}
    \end{subfigure}
    \hfill\vspace{7mm}
    \begin{subfigure}{0.48\textwidth}
        \centering
        \includegraphics[width=.8\linewidth]{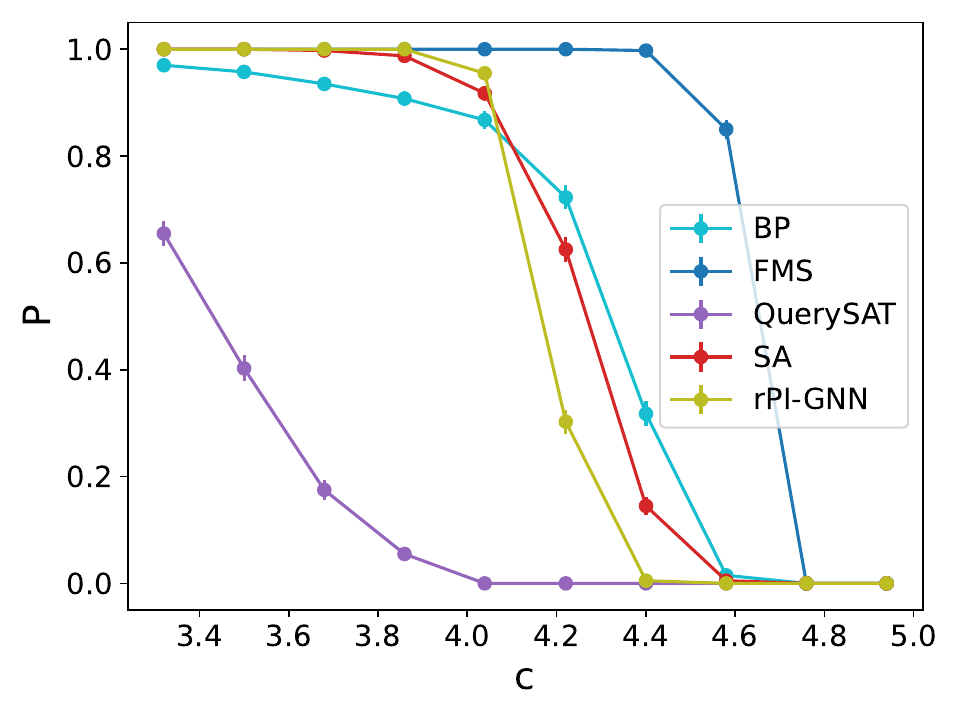}
        \caption{3-col $N=1024$}
        \label{fig:3col1024}
    \end{subfigure}

    \begin{subfigure}{0.48\textwidth}
        \centering
        \includegraphics[width=.8\linewidth]{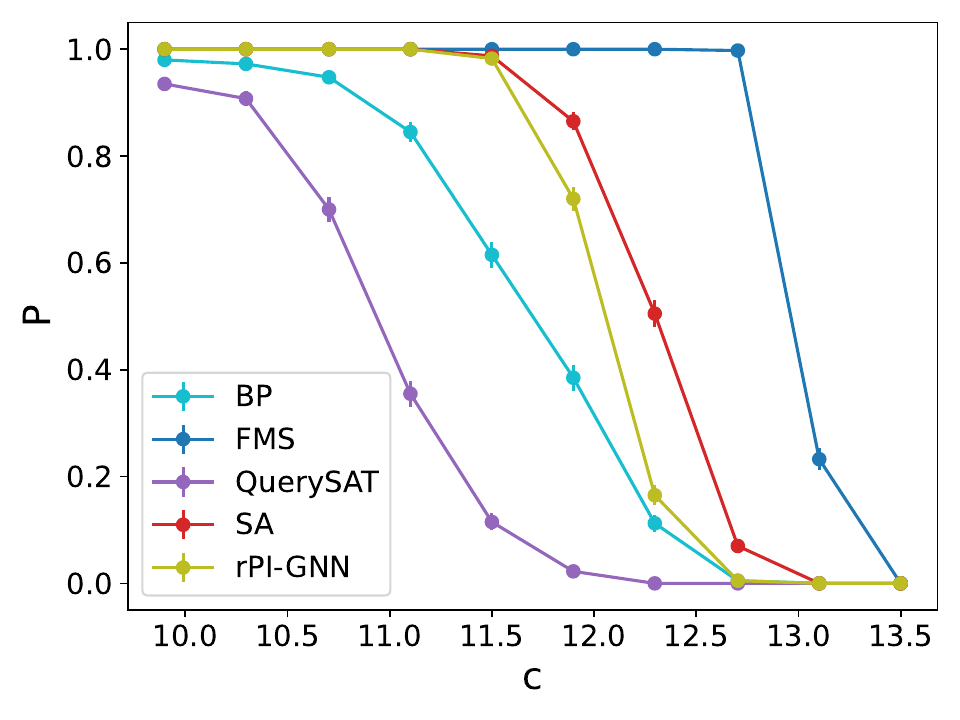}
        \caption{5-col $N=256$}
        \label{fig:5col256}
    \end{subfigure}
    \hfill\vspace{7mm}
    \begin{subfigure}{0.48\textwidth}
        \centering
        \includegraphics[width=.8\linewidth]{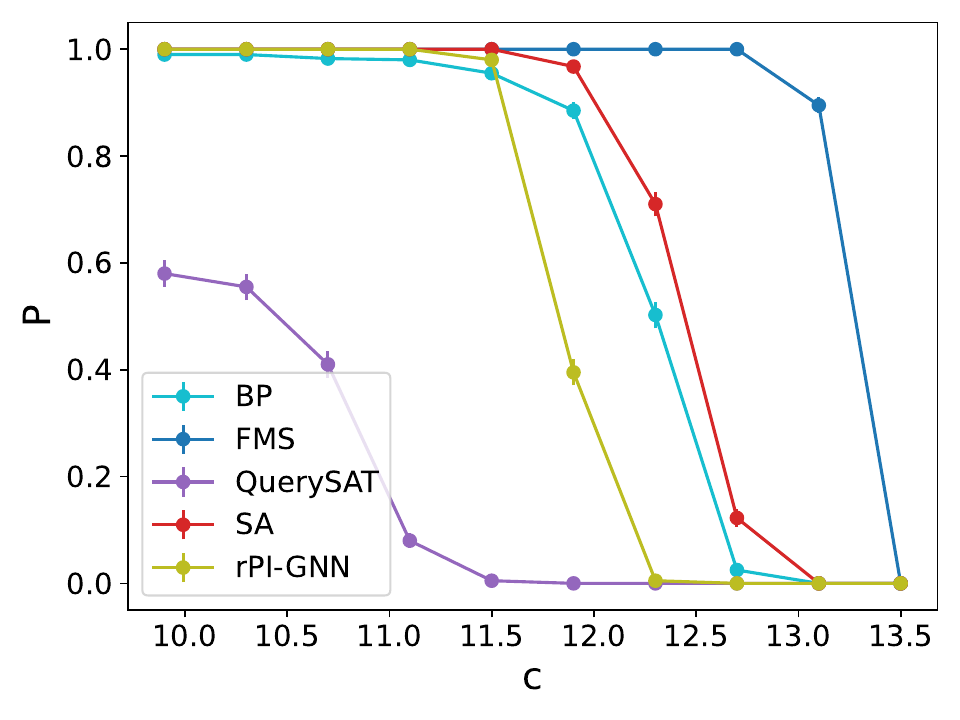}
        \caption{5-col $N=1024$}
        \label{fig:5col1024}
    \end{subfigure}
    \caption{\textbf{Probability of finding a satisfiable assignment using different algorithms} at fixed size ($N=256$ on the left and $N=1024$ on the right) for the $K$-SAT ($K=3,4$) and $q$-col ($q=3,5$) problems. We focus on two sizes: $N=256$ is the largest size in the training set, and $N=1024$, which is out-of-distribution, to check the generalization power.}
    \label{Fig:ProbabilityofSolution}
\end{figure}

\subsection{Comparisons of performances of different algorithms at fixed size}

In Fig.~\ref{Fig:ProbabilityofSolution} we look at the probability of finding a satisfiable assignment using different algorithms at a fixed size $N$ of the problem, changing the parameter $\alpha$ ($c$) for the $K$-SAT ($q$-col) problem with $K=3,4$ ($q=3,5$). We focus on two sizes: $N=256$, the largest in the training set, and $N=1024$, for which the GNNs are used out-of-distribution, to check their generalization power.
The best algorithm for both SAT and col problems is FMS. GNNs can reach good performances for $N=256$ in the $3$-SAT and $3$-col problems; however, their performances deteriorate for $N=1024$, because they are used out-of-distribution. They also rapidly deteriorate when applied to hard problems such as the $4$-SAT and the $5$-col for which classic algorithms are the most competitive ones.

In Table \ref{Tab:score}, we list the score reached by each algorithm, defined as the percentage of solved instances over the satisfiable ones. The considered instances are all the ones in the testing dataset, in the whole range of $\alpha$ ($c$) for the $K$-SAT ($q$-col) problem with $K=3,4$ ($q=3,5$) for all sizes up to $N=256$. The score computed on higher $N$ is reported in the Supplementary Information. The score provides a glance at algorithms' performances, but there are many other important features to consider: the scaling of running times with problem size, the power of generalization out-of-distribution, and the algorithmic threshold that we will analyze in the following sections.

In Table \ref{Tab:score}, we also list the residual energy for all algorithms, defined as the number of non-satisfied constraints divided by the number of constraints $M$, averaged over instances in the test set for which the algorithm did not find a solution. This is a measure complementary to the score. Let's look at MP algorithms as an example: they are good algorithms in terms of the score, but once they stop working, they return a mostly random configuration, thus resulting in a high residual energy.

\begin{table}
\begin{center}
\begin{minipage}{0.502\textwidth}
\centering
\resizebox{\textwidth}{!}{
  \begin{tabular}{ | c|| c | c | c | c |}
  \hline
    \multicolumn{1}{|c||}{\textbf{SAT}} & $K=3$; S $\uparrow$ & $K=4$; S $\uparrow$ & $K=3$; RE $\downarrow$ & $K=4$; RE $\downarrow$ \\
    \hline
    \rule{0pt}{10pt}NeuroSAT & 84.48  & 47.52 & 0.0072 & 0.0028 \\ \hline
    \rule{0pt}{10pt}QuerySAT & 92.38 & 66.57 & 0.0068 & \textbf{0.0027} \\ \hline 
    \specialrule{1.5pt}{0pt}{0pt}  
    \rule{0pt}{10pt}FMS &   \textbf{99.98} & \textbf{95.15} & \textbf{0.0061} & 0.0032
 \\ \hline 
    \rule{0pt}{10pt}SA & 98.75 & 82.61 & 0.0063 & \textbf{0.0027}
\\ \hline
\rule{0pt}{10pt}SP + dec. & 82.28 & 60.02 &  - & - \\ \hline
    
    \end{tabular}
    }
\end{minipage}
\hfill
\begin{minipage}{0.475\textwidth}
\centering
\resizebox{\textwidth}{!}{
  \begin{tabular}{ | c || c | c | c | c | }
  \hline
     \multicolumn{1}{|c||}{\textbf{col}}  & $q=3$; S $\uparrow$ & $q=5$; S $\uparrow$ & $q=3$; RE $\downarrow$ & $q=5$; RE $\downarrow$ \\
    \hline
    \rule{0pt}{10pt}rPI-GNN   & 92.49 & 68.68 & 0.0190 & 0.0131 \\ \hline 
    \rule{0pt}{10pt}QuerySAT  & 63.15 & 39.33 & 0.0229 & 0.0256\\ 
    \specialrule{1.5pt}{0pt}{0pt}  
 \rule{0pt}{10pt}FMS  & \textbf{99.96} &  \textbf{80.73}& \textbf{0.0196} & \textbf{0.0178} \\ \hline 
    \rule{0pt}{10pt}SA   & 90.25  & 74.54 & 0.0185 & 0.0232
\\ \hline
    \rule{0pt}{10pt}BP + dec. & 78.47 & 45.19& 0.0652 & 0.0638
    \\ \hline
    \end{tabular}
    }
\end{minipage}
\end{center}
\caption{\textbf{Score (S) and Residual Energy (RE) reached by each algorithm}. The Score is the ratio of solved instances over the satisfiable ones in the test dataset ($N \le 256$ and any $\alpha$ or $c$ values). An instance is considered satisfiable if at least one of the analyzed algorithms found a solution (see Sec. \ref{sec:Score}). The ratio of sat instances $\text{n}_{\text{sat}}/\text{n}_{\text{tot}}$ are 28656/42000 (3-SAT), 19806/21000 (4-SAT), 11750/20000 (3-col), and 9184/20000 (5-col). RE is the average of the intensive energy computed on instances for which the algorithm is not finding a solution. 
The running steps are $1000N$ for SA, $100N$ for FMS in the $K$-SAT problem, and $1000N$ for SA, $625N$ for FMS in the $q$-col problem, with each step involving $N$ attempted single-variable updates. The number of recurrent iterations is $2N$ for NeuroSAT and QuerySAT. Best results are in bold. RE values for the SP algorithm are missing because SP sometimes does not converge on hard instances, and in this case it does not output a configuration of the variables (see Sec. \ref{subsubsec:SP} for details).}
\label{Tab:score}
\end{table}

\subsection{Extracting algorithmic thresholds}
\label{sec:AlgThresh}

In analogy to the satisfiability threshold that concentrates in the large $N$ limit, for each algorithm and problem, an algorithmic threshold, $\alpha_\text{alg}$ or $c_{alg}$, can be defined such that in the large size limit only for $\alpha<\alpha_\text{alg}$ the chosen algorithm finds a solution with high probability for a typical instance of the random problem.
For the first time, we attempt to estimate algorithmic thresholds also for GNNS.

\begin{figure}  
\centering
    \begin{subfigure}{0.32\textwidth}
        \centering
        \includegraphics[width=\linewidth]{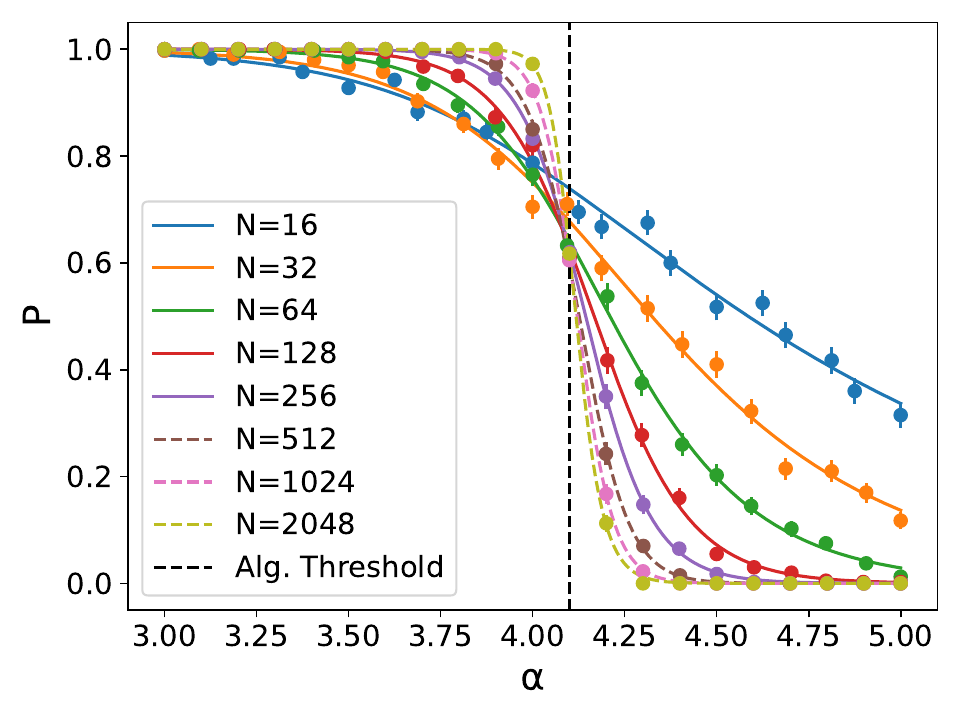}
        \caption{QuerySAT (3-SAT)}
        \label{fig:query_3sat_thresh}
    \end{subfigure}
    \hfill
     \begin{subfigure}{0.32\textwidth}
        \centering
        \includegraphics[width=\linewidth]{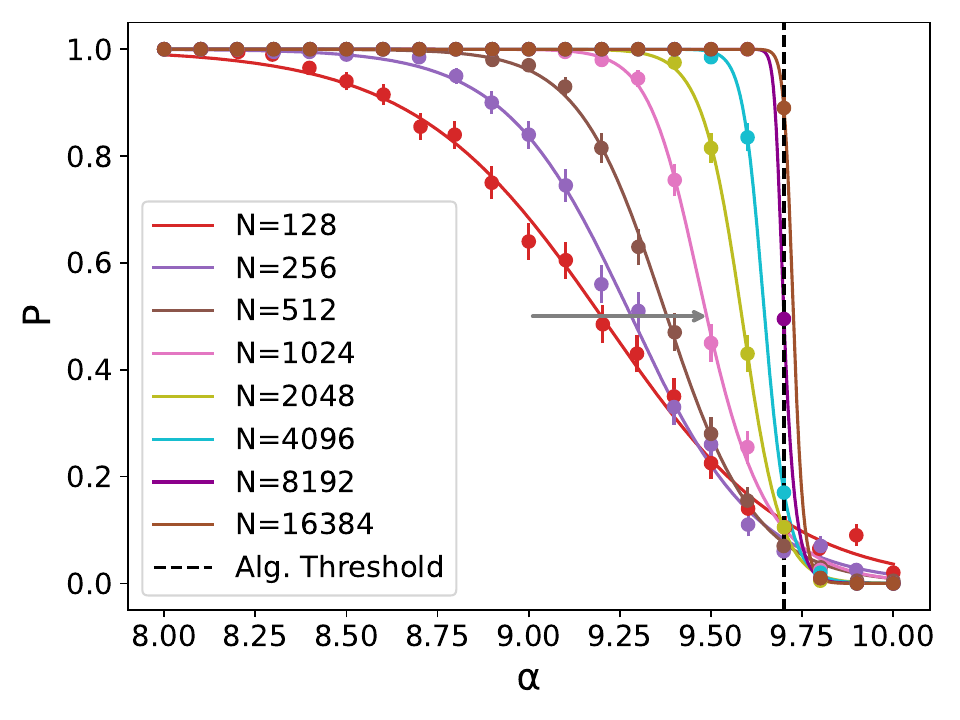}
        \caption{SP (4-SAT)}
        \label{fig:sp_4sat_thresh}
    \end{subfigure}
    \hfill
    \begin{subfigure}{0.32\textwidth}
        \centering
        \includegraphics[width=\linewidth]{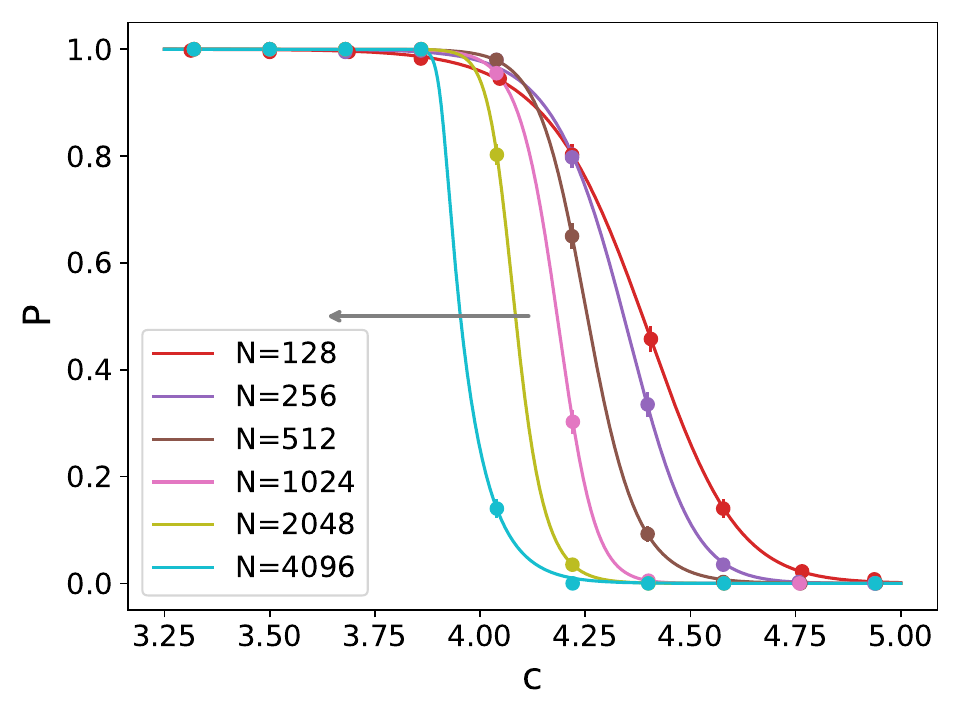}
        \caption{rPI-GNN (3-col)}
        \label{fig:rPIGNN_sub2}
    \end{subfigure}
    \caption{\textbf{Estimation of the algorithmic threshold in the large size limit} for different algorithms and problems, indicated in the label under the panel. The left panel shows the easy situation, when crossings of curves for different sizes take place at the same point (apart from small finite-size corrections), and the crossing point corresponds to the algorithmic threshold in the large $N$ limit. Instead, in the central and the right panels, we report more difficult situations, where the crossings are not present or happen very close to boundaries ($P=0$ or $P=1$). In the latter cases, the curves measured at finite $N$ provide a bound to the algorithmic threshold in the large $N$ limit. Data in the central panel provide a lower bound, while those in the right panel provide an upper bound. }
    \label{Fig:algorithmicThreshold}
\end{figure}

For this purpose, we use standard procedures from statistical physics: looking at the sat probability as a function of $\alpha$ (or $c$) for different sizes, and supposing that this probability in the infinite-size limit becomes a step function with a jump in $\alpha_\text{alg}$ (or $c_{alg}$), we estimate the algorithmic threshold from the crossing of the curves corresponding to different sizes. An alternative determination of the algorithmic threshold based on the crossing of residual energy, that gives analogous results, is reported in the Supplementary Information. Different situations could arise, all shown in Fig.~\ref{Fig:algorithmicThreshold}.
The easy situation is when, for $N$ large enough, the curves with different $N$ values cross at the same point (apart from small finite size corrections), and this identifies a well-defined algorithmic threshold in the large $N$ limit (see left panel in Fig.~\ref{Fig:algorithmicThreshold}).
The difficult situations arise when the crossings happen very close to one of the boundaries (see the central panel in Fig.~\ref{Fig:algorithmicThreshold}, with crossing very close to $P=0$) or are completely absent from data (see right panel in Fig.~\ref{Fig:algorithmicThreshold}). In these cases, the curves that move rightwards (resp.~leftwards) increasing $N$ as in the central (resp.~right) panel just provide a lower (resp.~upper) bound to the algorithmic threshold in the large $N$ limit.
In particular, performances of SP solving random 4-SAT improve increasing $N$ and an estimate for the algorithm threshold is still possible (vertical dashed line), while performances of rPI-GNN in solving random 5-col degrades a lot increasing $N$, and the estimate is impossible in this case.

\begin{table}
\begin{center}
  \begin{tabular}{ | c || c | c | }
    \hline
    \textbf{SAT} & $K=3$ & $K=4$ \\
    \hline
    NeuroSAT & 3.8 & $\lesssim$ 8.1 \\
    \hline
    QuerySAT & 4.1 & 9.1 \\
    \specialrule{1.1pt}{0pt}{0pt} 
    FMS & \textbf{4.2}  & \textbf{9.8} \\
    \hline 
    SA & 4.1 & 9.5 \\
    \hline
    SP + dec. & \textbf{4.2} & 9.7 \\
    \hline
  \end{tabular}
\hspace{1cm}
  \begin{tabular}{ | c || c | c | }
    \hline
    \textbf{col} & $q=3$ & $q=5$ \\
    \hline
    rPI-GNN & $\lesssim$ 4.0 & $\lesssim$ 11.4 \\
    \hline
    QuerySAT & $\lesssim$ 3.48 & $\lesssim$ 10 \\
    \specialrule{1.1pt}{0pt}{0pt} 
    FMS & \textbf{4.4} & \textbf{13.0} \\
    \hline
    SA & 4.2 & 12.6 \\
    \hline
    BP + dec.& $\gtrsim$ \textbf{4.4} & $\gtrsim$ 12.4 \\
    \hline
  \end{tabular}
\end{center}
\caption{\textbf{Algorithmic threshold for different algorithms} By definition, the algorithmic threshold is computed taking the limit of large $N$. In this limit, Neural Solvers behave much worse than classical algorithms, especially in the hard 1RSB problems ($4$-SAT and $5$-col). FMS is the algorithm that has the best algorithmic thresholds.}
\label{Tab:alphaAlg}
\end{table}

Performing such analysis for all the tested algorithms on the benchmark instances, we obtain the values of $\alpha_\text{alg}$ reported in Table \ref{Tab:alphaAlg}.
We do not pretend to extract algorithmic thresholds with extreme precision; the interested reader can find more accurate estimates for the classic algorithms in the referenced papers. We just put two significant digits because they are enough to compare the different algorithms. The main message of Table~\ref{Tab:alphaAlg} is that, even if GNN can solve efficiently instances of small sizes, when moving towards larger sizes their performances severely degrade. As a consequence, their algorithmic threshold is much lower than the ones of classical algorithms. This is particularly true in the harder cases $K=4$ and $q=5$, while differences are smaller in the easier cases $K=3$ and $q=3$. Thus using only the latter for a comparison (as it has been done in the literature) is not very meaningful.

\subsection{Comparison of running times for different algorithms}
\label{sec:runningtimes}

\begin{table}
\begin{center}
  \begin{tabular}{ | l || l | l | }
    \hline
    \textbf{SAT} & $K=3$ & $K=4$ \\
    \hline
    NeuroSAT & 14.57 & 24.91 \\
    \hline
    QuerySAT &134.43 & 322.53 \\
    \specialrule{1.1pt}{0pt}{0pt} 
    FMS & 1.83 & 3.80 \\
    \hline 
    SA & \textbf{1.82} & 13.04 \\
    \hline
    SP + dec. & 2.61 & \textbf{3.06} \\
    \hline
  \end{tabular}
\hspace{1cm}
  \begin{tabular}{ | l || l | l | }
    \hline
    \textbf{col} & $q=3$ & $q=5$ \\
    \hline
    rPI-GNN & 61.44 & 107.91 \\
    \hline
    QuerySAT & 4.95 & 24.89 \\
    \specialrule{1.1pt}{0pt}{0pt} 
    FMS & \textbf{0.006} & \textbf{0.02} \\
    \hline
    SA & 65.62 & 88.06 \\
    \hline
    BP + dec. & 56.70 & 441.50 \\
    \hline
  \end{tabular}
\caption{\textbf{Average running times} (in seconds) for different algorithms over $400$ instances with $N=1024$ variables. For neural solvers, we report the testing time. We fix $\alpha=4.1$ for 3-SAT, $\alpha=9.4$ for 4-SAT, $c=3.68$ for 3-col and $c=11.1$ 5-col. For $K$-SAT, we ran the algorithms using Lenovo Hardware AMD Ryzen Threadripper PRO 5975WX 32-Cores with 251 GB of RAM. For $q$-col, we ran the rPI-GNN using the NVIDIA Tesla V100 SXM2 32G GPU and QuerySAT on NVIDIA TITAN RTX GPU, while all other algorithms were run on a single CPU Intel Xeon Gold 6248 2.5G.
}
\label{Tab:actualTimes}
\end{center}
\end{table}

Finally, we have measured the (average) wall-clock running time of the different algorithms and report them in Table \ref{Tab:actualTimes}.

It is worth stressing that neural solvers like NeuroSAT and QuerySAT require a time for training which is much larger than the time to solve an instance once trained. For NeuroSAT, this training time is approximately 37min/epoch on 3SAT and 19 min/epoch on 4SAT, while for QuerySAT it is approximately 27min/epoch on 3SAT and 17 min/epoch on 4SAT. The difference in time is motivated by the fact that, as previously mentioned in Section \ref{sec:main-datasets}, the 4SAT dataset is half the size of the 3SAT dataset, and the increase in $k$ can be easily bypassed through available GPU memory. Moreover, rPI-GNN (as its original version PI-GNN) requires a new training for each sample: for this reason, its computing time is much higher than other standard neural solvers. At test-time, we fix a number of message-passing iterations of $t=2N$ for all neural solvers.

Classical heuristics algorithms do not need any training, and their running times are essentially fixed by the schedule we have chosen for the maximum number of iterations: $t=100 N$ in SAT problems and $t=625 N$ in col problems for FMS, and $t=1000N$ for SA, both in SAT and col problems.
However, we noticed a difference between FMS and SA: while SA finds the solution (or the optimal configuration) close to the end of the run, FMS may find a solution much earlier (this is the reason why the actual running time reported in Table \ref{Tab:actualTimes} is so much shorter than the others).

\section{Discussion}

In this paper, we have introduced a new benchmark based on two well-known CSPs, $K$-SAT and $q$-col, with the following key features:
\begin{itemize}
    \item Controlled complexity variation through parameters $K$, $\alpha$ ($K$-SAT), $q$, $c$ ($q$-col) and different problem sizes $N$;
    \item Consistent training/test splits with reproducible generation;
    \item Compatibility with standard CSP formats (e.g., DIMACS CNF for SAT).
\end{itemize}

This structured benchmark enables systematic evaluation and comparison of CSP solvers across varying complexity levels and generalization demands. Using this benchmark, we compared classical algorithms with newer GNN-based solvers. Our experiments revealed that supervised GNNs underperform, leading us to focus on unsupervised approaches.

Inspired by recent literature, we trained neural solvers for a fixed number of iterations while we let the testing time scaling linearly with the problem size $N$, confirming that this is necessary to prevent performance degradation as $N$ grows (a requirement also observed in classical algorithms). 
Please note that fundamental limitations for running GNN on random instances hold when the depth of GNN does not scale with graph size, from the presence of the overlap gap property phase transition \cite{gamarnik2023barriers}. However, when GNN are used for a time that scales with $N$, as in this work, the hypothesis of ref. \cite{gamarnik2023barriers} are no more valid.

Our numerical tests show that, while neural solvers perform comparably to classical methods for easier problems (3-SAT and 3-col), their performance declines significantly for harder problems (4-SAT and 5-col). This aligns with known geometric transitions in the solution space that hinder algorithmic efficiency. Classical algorithms still deal better with this obstacle. When developing a new NN algorithm, it is thus crucial to test it on these harder problems.

Another key point is the scaling of performances with $N$. Notably, the PI-GNNs from ref.~\cite{Schuetz_PIGNNcol_2022} do not employ traditional training; instead, their weights are optimized per-instance, resembling classical optimizers. While this approach has merits, it sacrifices one of GNNs’ key advantages: fast inference after a single training phase. Consequently, PI-GNNs face scalability challenges at large $N$. On the other side, NeuroSAT and QuerySAT can be trained in a reasonable time just for small sizes ($N\leq 256$), and their performances clearly degrade when used in testing out-of-distribution.
In contrast, classical algorithms exhibit stable performance regardless of $N$: this is a key feature that newly designed NN should obtain. 

Given the substantial room for improvement in GNN-based CSP solving, we include large-$N$ instances (up to $N = 16384$) in our released dataset, along with results from FMS and SP (the best-performing classical solvers in our study). We hope these resources will encourage future work toward neural solvers that match or surpass classical performance.


\section{Methods}
\label{sec:methods}

\subsection{The random $K$-SAT and $q$-col problems}
\label{sec:Ksat_qcol}
The random-$K$-SAT problem, the prototype of CSPs, asks whether there exists an assignment of $N$ Boolean variables that satisfies at the same time $M$ given clauses (constraints). Each instance of the problem is specified by a formula consisting of a conjunction of $M$ clauses, where each clause is a disjunction of literals i.e., a variable or its negation. In the $K$-SAT problem, every clause contains exactly $K$ literals:  $(x_1\lor  \neg x_3\lor x_7)$ is an example of a clause for a 3-SAT problem, where $x_i$, $i=1,2...,N$ are the $N$ Boolean variables, and we indicate with $\neg x_i$ their negation. The clustering, condensation and satisfiability threshold values for $K=3$ and $K=4$ are respectively: $\alpha_d(3)=\alpha_c(3)=3.86$, $\alpha_s(3)=4.267$; $\alpha_d(4)=9.38$, $\alpha_c(4)=9.547$, $\alpha_s(4)=9.931$ \cite{Montanari_2008,MertensKSAT2006}. 
One can introduce the following cost or energy function associated with the $K$-SAT problem: 
\begin{equation}
E^{K-\text{SAT}}(\{x\})=\sum_{a=1}^{M}\prod_{i\in\partial a}\left(\frac{1+J_{ia}\cdot(-1)^{x_{i}}}{2}\right),
\label{eq:E_K-SAT}
\end{equation}
where with $i\in\partial a$ we indicate all the nodes $i$ that enter in the $a$-th clause, $J_{ia}=1$ if $x_{i}$ enters the clause $a$ as it is and
$J_{ia}=-1$ if $x_{i}$ enters the clause $a$ negated.

Many optimization algorithms described in the following section will try to minimize this energy function.

The random $q$-col problem asks for an assignment of colors to the $N$ nodes of an undirected random graph (regular or Erd\H{o}s-Renyi) with finite (average) degree $c$. Each node must be assigned one of $q$ possible colors, under the constraint that no two neighboring nodes share the same color.
The minimum number of colors $q_{min}$ needed to properly color a graph is called the chromatic number. For any connectivity larger than the satisfiability one $c>c_S(q)$, the chromatic number is greater than $q$ for any typical instance. The exact values of the thresholds for a Poissonian random graph, at $q=3$ and $q=5$ are: $c_d(3)=c_c(3)=4$, $c_S(3)=4.687$, $c_d(5)=12.837$, $c_c(5)=13.23$, $c_S(5)=13.669$ \cite{zdeborova2007phase}. 
Similarly to the $K$-SAT problem, we can introduce an energy, or cost function, also for the $q$-col problem:
\begin{equation}
    E^{q-\text{col}}(\{x\}) = \sum_{(i,j)\in E} \delta_{x_i x_j},
\label{eq:E_qcol}
\end{equation}
where $x_i$ indicates the color assigned to node $i$ and can take integer values in the interval $[1,q]$, and the sum runs over the set $E$ of all the edges of the graph.
Again, many algorithms will try to minimize this energy function when trying to solve the $q$-col problem. 

\subsection{Dataset}

\subsubsection{$K$-SAT}
Random $K$-SAT instances are generated using a custom implementation that constructs formulas in Conjunctive Normal Form (CNF). Each CNF instance is defined by $N$ Boolean variables and $M = \mathrm{round}(\alpha N)$ clauses, where $\alpha$ is the clause-to-variable ratio. Each clause contains $K$ literals, generated by sampling $K$ distinct variables uniformly at random from the $N$ available. Each selected variable is independently negated with probability $0.5$, ensuring an unbiased distribution of positive and negative literals.
We note that this method of generation allows duplicate clauses in a formula. However, in the large $N$ limit, if $M$ is proportional to $N$, the probability of having a duplicated clause goes to zero.

For the $K$-SAT problem we built a training and a testing set.
The values of $\alpha$ for the instances in both the training and the test set are $\alpha \in [3,5]$ with $\Delta \alpha=0.1$ for $3$-SAT and $\alpha \in [8,10]$ with $\Delta \alpha=0.1$  for $4$-SAT.
In the training set, for each of the 21 values of $\alpha$ when $K=3$ (respectively $K=4$), we generated 1600 (respectively 800) instances for each value of $N \in \{16, 32, 64, 128, 256\}$. 
To ensure reproducibility, we use stable pseudo-random number generators seeded with a combination of a base seed and the variable size $N$.

For each instance on the training set we also include the solution given by a SAT solver (CaDiCaL~\cite{BiereFallerFazekasFleuryFroleyksCADICAL4}).
The GNN analyzed in this paper are trained on these training sets, either in a supervised or unsupervised way.

In the test set for each of the 21 values of $\alpha$ when $K=3$ (respectively $K=4$), we generated 400 (respectively 200) instances for each value of $N \in \{16, 32, 64, 128, 256\}$: these instances are the one used to generate Figg. \ref{fig:neurosat-linscaling-iter}, \ref{Fig:ProbabilityofSolution} and to compute score and residual energy of Table \ref{Tab:score}. Instructions on how to obtain the training and the test datasets with sizes up to $N=256$ are provided at GitHub repository of the project. In the GitHub repository, we also added the results of the evaluation for all the analyzed algorithms on this test dataset.

However, as explained in sec. \ref{sec:AlgThresh}, it is important to test the effectiveness of an algorithm in the large size limit. For this reason, in the test set we also included instances with the same chosen values of $\alpha$ and with $N \in \{512, 1024, 2048, 4096, 8192, 16384\}$ for both $K=3$ and $K=4$. We will call these instances the Out-of-distribution (OOD) instances. All instances can be built with the released code in the GitHub repository.
We used these OOD instances (or part of it, for some algorithms) to evaluate the algorithmic thresholds in Tab. \ref{Tab:alphaAlg} and Fig. \ref{Fig:algorithmicThreshold}. We also evaluated the score for the chosen algorithms on part of this dataset, reporting it in the Supplementary information.
These OOD instances are mainly intended as a challenge for the community: a good newly proposed GNN should be able to reach results comparable with (or even better than) what is obtained by classical algorithms on this dataset. For the whole OOD test set, we include in the repository the results of FMS, the best algorithm we analyzed so far, as a comparison for future developed algorithms.

\subsubsection{$q$-coloring}
To build the $q$-col training and testing datasets, we follow the statistical physics community, i.e., we build random graphs with some essential features \cite{zdeborova2007phase}. The ensemble of graphs considered in this work is defined by the statistics of the average connectivity $c=2\frac{M}{N}$, $M$ being the number of undirected edges and $N$ the number of nodes. Such statistics should not depend on $N$. The connectivities considered do not increase with $N$, so that $c=\mathcal{O}(1)$. The graphs are sparse, and they possess a tree-like local structure. More precisely,  we considered the case of Erdős–Rényi ensemble with fixed $M$, $G(N,M)$: $M$ links are assigned uniformly at random between couples of nodes, without repetitions. After the edges are assigned, the connectivity is exactly $c=2\frac{M}{N}$. We remark that there are $\binom{\binom{N}{2}}{M}$ possible graphs that can be constructed, and we extract graphs from this ensemble with uniform probability. Our analysis is performed using the value of $c$ as the control parameter.

As in the $K$-SAT case, we create a training dataset with sizes up to $N=256$ while the testing dataset contains also higher OOD sizes.

For both the training and test set we include graphs with values of average connectivity of nodes $c \in [3.32, 4.94]$ with $\Delta c=0.18$ for $3$-col and $c \in [9.9, 13.5]$ with $\Delta c=0.4$ for $5$-col. In the training set for each value of $c$ when $q=3$ (respectively $q=5$), we generated 1200 instances for each value of $N \in \{16, 32, 64, 128, 256\}$. We do not furnish solutions for this training set, because we only used unsupervised learning protocols.
For the test set, for each value of $c$ when $q=3$ (respectively $q=5$), we generated 400 instances for each value of in-distribution $N \in \{16, 32, 64, 128, 256\}$ and for each OOD $N \in \{512, 1024, 2048, 4096, 8192, 16384\}$, which were used to evaluate algorithm performance on large-scale problems. All instances can be built with the released code.

As already explained, we focused on the assignment problem. Some previous works also focused on the decision problem, finding some nonphysical predictions, with GNN predicting colorable graphs with a given $q$ while the chromatic number of that particular instance was larger than $q$ \cite{lemos2019graph}. To avoid these contradicting outputs, the performances of the different algorithms on the dataset will be compared only on the assignment problem, both for $K$-SAT and $q$-col.

\subsection{Classical Heuristics Algorithms} \label{sec:classicAlg}

\subsubsection{Simulated Annealing}

Simulated Annealing (SA) is a standard way to use Monte Carlo Markov Chains (MCMC) to solve combinatorial optimization problems \cite{KirkpatrickSA1983}. With more than 40 years of history, SA has many variants \cite{EgleseSimAnn1990, GuilmeauSimAnn2021} and has been applied to several combinatorial optimization problems \cite{KirkpatrickSA1983, zdeborova2010generalization, krzakala2013performance, Budzynski_2019, mariachiara_indset_PRE_2019, mariachiara_SA_col_2023}.

The goal of a generic MCMC is to sample configurations $\underline{x}$ of the variables in the problem following the Gibbs measure $P(\underline{x}) = e^{-\beta E(\underline{x})} / Z$. Here, $Z$ is a normalization factor and $\beta$ is a nonnegative parameter. In physics, $Z$ is known as the partition function, and $\beta$ is the inverse of temperature $T$ ($\beta=1/T$). In the scenario of combinatorial optimization, $E(\underline{x})$ is the cost function to minimize in order to find solutions of the problem. The corresponding energy functions for the $K$-SAT and the $q$-col problems are reported in eqs. \eqref{eq:E_K-SAT}, \eqref{eq:E_qcol}. 

The MCMC is at the core of the SA algorithm in combinatorial optimization for a simple reason: in the limit $\beta \to \infty$ (zero temperature), solutions become exponentially dominant minima of the Gibbs measure. Thus, the correct sampling at $T=0$ would be equivalent to a uniform sampling of solutions. However, this is not always an easy task. Most cost functions in relevant combinatorial optimization problems display rugged landscapes that would trap a zero-temperature MCMC in local minima before accessing actual solutions.

To overcome this problem, it is convenient to accept a certain level of noise (finite $\beta$ or $T>0$) at the beginning of the landscape's exploration. The key in SA is the selection of a schedule to decrease the temperature, \textit{i.e.}, the noise, from an initial level $T_0$ to a final value $T_n$ where one hopes to find solutions. The so-called \emph{cooling schedule} consists of a sequence $T_0 > T_1 > \ldots > T_n$. At each temperature $T_k$, one or more steps of the regular MCMC are performed. When consecutive temperatures in the schedule are close enough, one aims to always sample from a Gibbs measure that gradually concentrates toward solutions. 

In this work, we used a cooling schedule with constant temperature steps. Although this traditional design \cite{KirkpatrickSA1983} might not be optimal, we preferred it for its simplicity while expecting the algorithmic transitions not to strongly depend on our choice.

The reader might find the following pseudo-code useful to complement our description:
 \begin{algorithm}[H]
\begin{algorithmic}[1]
    \State{Set a random initial $\underline{x}_0$}
    \State{Choose a number of steps $n$}
    \State{Set a \textit{cooling schedule} $T_0 > T_1 > \ldots > T_n$, with $T_k = T_0(1-k/n)$}
    \For{$k=0, \ldots, n$}
    \State{Take $N$ Metropolis steps at $T_k$}
    \EndFor \\
    \Return{Final cost $E_f$ and final configuration $\underline{x}_n$}
  \end{algorithmic}
  \caption{Simulated Annealing}
 \label{alg:SA}
\end{algorithm}

The Metropolis step \cite{Metropolis1953} is defined as follows: when the algorithm is visiting a configuration $\underline{x}$, a site $i$ is randomly selected and a new $x'_i \neq x_i$ is proposed uniformly at random among all possible choices. In particular, in the $K$-SAT problem, the proposed move changes the variable to the other possible value, while in the $q$-col problem a new color among the $q-1$ different ones is randomly proposed. Then, this change is accepted with probability:

\begin{equation}
r(\underline{x} \to \underline{x}') = \begin{cases}
                           1 \quad\qquad\qquad\quad \text{if} \:\: \Delta E(\underline{x} \to \underline{x}') \leq 0 \\
                           e^{-\beta \, \Delta E(\underline{x} \to \underline{x}')}\quad \text{if} \:\: \Delta E(\underline{x} \to \underline{x}') > 0
                          \end{cases}
 \label{eq:Metropolis_rates}
\end{equation}
where $\Delta E(\underline{x} \to \underline{x}') = E(\underline{x}') - E(\underline{x})$ and $\underline{x}'$ is identical to the configuration $\underline{x}$ in all sites but the $i$-th.

Notice that by running the SA for a fixed value of $n$ we get a linear-time algorithm since we perform $N$ simple Metropolis steps at each temperature. Following the ideas in Ref. \cite{angelini2025algorithmic}, we explored several superlinear time scalings to enhance the performance of SA. We used quadratic scalings where one increases $n$ linearly with the system size $N$. The resulting real running times still make SA faster or at least competitive with the rest of the algorithms in this work.
\subsubsection{Focused Metropolis Search} \label{subsubsec:FMS}

Focused Metropolis Search (FMS) is a member of the family of stochastic local search algorithms tailored for random $K$-SAT. Introduced in Ref. \cite{Seitz_2005}, FMS achieves performances similar to the best versions of the well-known WalkSAT algorithm \cite{AurellNIPS2004WalkSAT, Seitz_2005}, solving 4-SAT instances in linear time even beyond the condensation transition \cite{AlavaPNAS2008}.

Once fed with an initial assignment of the variables, the algorithm proceeds as follows. First, it selects an unsatisfied clause uniformly at random among all unsatisfied clauses produced by the current assignment. Second, it chooses a variable $x_i$ inside that clause also uniformly at random. FMS evaluates the possibility of flipping $x_i$ using the well-known Metropolis rule: If the flip diminishes the total energy $E$, it is accepted; if it increases $E$ by an amount $\Delta E$, it is accepted with probability $\eta^{\Delta E}$. 
As for the Simulated Annealing algorithm, the corresponding energy functions for the $K$-SAT and the $q$-col problems are the ones reported in eqs. \eqref{eq:E_K-SAT}, \eqref{eq:E_qcol}.

After looking back at the Metropolis rates as presented in eq. \eqref{eq:Metropolis_rates}, one sees that $\eta=e^{-\beta}$ is another way to parameterize the level of noise in the dynamics. When $\eta=0$ (which corresponds to $\beta \to \infty$), we have a greedy dynamics that never goes up in energy. The other extreme $\eta=1$ corresponds to what is known as random WalkSAT \cite{PapadimitrouWalkSAT}, where all flips are accepted. 

The algorithmic parameter $\eta$ must be tuned to achieve optimal results while guaranteeing a balance between exploration and exploitation moves. Following the literature, we choose the values $\eta=0.37$ for 3-SAT \cite{Seitz_2005} and $\eta=0.293$ for 4-SAT \cite{AlavaPNAS2008}. As for the running times, we selected quadratic scalings $t = A N^{2}$ with different coefficients $A$, obtaining outstanding performances when $A$ is large enough. In $K$-SAT instances, we used $A=100$, guaranteeing that FMS's maximum wall-clock time is comparable to that of the other classical heuristics for $K$-SAT in Table \ref{Tab:score}, taking SA with $t=1000N^{2}$ as reference.

FMS is easily adaptable to other combinatorial optimization problems. Here, we also introduce FMS for the $q$-col of random graphs in what is, to our knowledge, its first implementation for a problem other than $K$-SAT. The procedure is essentially the same. First, we choose an unsatisfied edge, connecting two nodes with the same color, uniformly at random. Then, one of these two nodes is randomly selected, and we propose a new color for it among all $q-1$ possibilities. The new color is accepted using the Metropolis rule with noise $\eta$, as before.

Being this the first implementation for $q$-col, there are no previous results on the optimal noise $\eta$ to use. After a non-exhaustive exploration, we choose $\eta=0.37$ for $q=3$ and $\eta=0.25$ for $q=5$. For the same reason as with $K$-SAT, we ran FMS for a time that scales with the system size as $t=625N^{2}$. This gives a maximum wall-clock time comparable to what we obtained applying SA to $q$-col with $t=1000N^{2}$.

To compute the scores in Table \ref{Tab:score}, we needed to identify as many satisfiable instances as we could. Since FMS is the best-performing algorithm in our testing datasets for $K$-SAT and for $q$-col, we ran it for longer times to enhance its capacity to solve instances, paying the price of impractically long wall-clock times. Its solutions with the time scaling $t=1000N^{2}$ are responsible for most of the instances we could identify as satisfiable.

\subsubsection{The Survey Propagation Algorithm for $K$-SAT} \label{subsubsec:SP}

The Survey Propagation algorithm (SP) is a heuristic message passing algorithm. It was developed by Mezard, Parisi, and Zecchina \cite{MezardSP} from the assumption of one-step replica symmetry breaking and the cavity method of spin glasses. SP works on the factor graph with the underlying CNF formula. While for small $N$, SP has convergence issues due to the presence of short loops in the graphs, for larger $N$, SP is conjectured to work better because cycles in the graph are of average length $O(\log N)$. Such large cycles allow, from a practical point of view, to consider large graphs as locally tree-like. The presence and absence of these loops, respectively, at small and large sizes, is exactly the reason for the quite small score reported in Table \ref{Tab:score}, computed on relatively small instances $N\leq 256$ and for the large algorithmic threshold reported in Table \ref{Tab:alphaAlg}, computed on large instance sizes. A detailed description of the Survey Propagation algorithm can be found in \cite{MezardSP, braunstein2005survey}. Here, we summarize it. 

Broadly speaking, SP exchanges messages between the $N$ variables and $M$ clauses in order to guess the value that each variable needs to be set to satisfy all clauses. More precisely, a message of SP, called a survey, passed from one function node $a$ to a variable node $i$ (connected by an edge) is a real number $\eta_{a\to i} \in [0, 1]$. The messages have a probabilistic interpretation under the assumption that SP runs over a tree-like factor graph, which is true in the large $N$ limit. In particular, the message $\eta_{a\to i}$ corresponds to the probability that the clause $a$ sends a warning to variable $i$, telling which value the variable $i$ should adopt to satisfy itself \cite{parisi2003probabilistic, ManevaSP}.

The iterative equations of SP are:
\begin{equation}
\label{algoSP}
\begin{split}
&s_{j\to a}^{\mp}=\left[1- \prod_{b \in \partial_{ja}^{\mp}} (1-\eta_{b\to j}) \right]\prod_{b \in \partial_{ja}^{\pm}} (1-\eta_{b\to j})\\
&s_{j\to a}^{0}=\left[\prod_{b \in \partial_{j} \setminus a} (1-\eta_{b\to j}) \right]\\
&\eta_{a\to i}=\prod_{j \in \partial_{a} \setminus i} \left[ \frac{s_{j\to a}^{-}}{s_{j\to a}^{-}+s_{j\to a}^{+}+s_{j\to a}^{0}}\right];
\end{split}
\end{equation}
where the symbol $\partial_a$ defines the set of variables nodes connected with the functional node $a$, i.e., the variable in clause $a$, and the symbol $\partial_i$ defines the set of functional nodes connected with the variable node $i$, i.e., the set of clauses where the literal $x_i$, or $\neg x_i$, appears. The cardinality of the set $\partial_i$ is the degree of a variable node $i$, i.e., the number of links connected to it, and is defined with $n_i$. The set $\partial_i$ is composed of two subsets, namely $\partial_i^+$, which contains the functional nodes where the variable node $i$ appears not negated, and $\partial_i^-$, which contains the functional nodes where the variable node $i$ appears negated. Obviously, the relation $\partial_i= \partial_i^+ \cup \partial_i^-$ holds.

With the symbol $\partial_{ia}^{+}$ (respectively $\partial_{ia}^{-}$) we define the set of functional nodes containing the variable node $i$, excluding the functional node $a$ itself, satisfied (respectively not satisfied) when the variable $i$ is assigned to satisfy clause $a$. In other words, if the variable $x_i$ is not negated in the clause $a$, then the $\partial_{ia}^{+}$ is the set of functional nodes containing the variable node $i$, excluding the functional node $a$ itself, where the variable $x_i$ appears not negated, while $\partial_{ia}^{-}$ is the set of functional nodes containing the variable node $i$, where the variable $x_i$ appears negated, $\neg x_i$. In contrast, if the variable $x_i$ is negated in the clause $a$, then the $\partial_{ia}^{+}$ is the set of functional nodes containing the variable node $i$, excluding the functional node $a$ itself, where the variable $x_i$ appears negated, while $\partial_{ia}^{-}$ is the set of functional nodes containing the variable node $i$, where the variable $x_i$ appears not negated.
 
SP is a local algorithm that takes as input a CNF formula of a random SAT Problem, and it performs a message-passing procedure to obtain convergence of the messages. More precisely, we are given a random initialization to all messages, and at each iteration, each message is updated following eq. \eqref{algoSP}. SP runs until all messages satisfy a convergence criterion: the iteration is halted at the first time $t^*$ when no message has changed by more than $\epsilon$, a fixed small number, over the last iteration. If this convergence criterion is not satisfied after $t_{max}$ iterations, SP stops and returns a failure output \cite{Marino_2021}. Once a convergence of all messages $\eta_{a \to i}$ is found, SP computes the marginals for each variable $i$:
\begin{equation}
\label{SID}
S_i^-=\frac{\pi_i^-(1-\pi_i^+)}{1-\pi_i^+\pi_i^-},\quad
S_i^+=\frac{\pi_i^+(1-\pi_i^-)}{1-\pi_i^+\pi_i^-},\quad
S_i^0=1-S_i^--S_i^+,
\end{equation}  
where:
\begin{equation}
\pi_i^{\pm}=1-\prod_{b\in \partial_i^{\pm}}(1-\eta_{b \to i}).
\end{equation}
The SP marginal $S_i^{+}$ ($S_i^{-}$) represents the probability that the variable $i$ must be forced to take the value $x_i=1$($x_i=0$), conditional on the fact that it does not receive a contradictory message, while $S_i^{0}$ provides the information that the variable $i$ is not forced to take a particular value. 

Once all the SP marginals have been computed, the decimation strategy can be applied. This subroutine is called Survey-Inspired Decimation (SID). Decimating a variable node $i$ means fixing the variable to $1$ or $0$ depending on the SP marginals, removing all satisfied functional nodes and the variable node $i$ from the factor graph, and removing all the literals from the clauses that have not been satisfied by the fixing. We select the variable node $i$ to decimate as the variable with the maximum bias $b_i=1-\min(S_i^-, S_i^+)$. Decimated the variable node $i$, the SP algorithm tries to find out a new state of convergence and uses decimation again until one of these three different outcomes appears: (i) a contradiction is found, then the algorithm returns exit failure; (ii) SP does not find a convergence, then the algorithm returns exit failure; (iii) all the messages converge to a trivial fixed point, i.e., all the messages are equal to $0$. 
In our analysis, the outcome (ii) has been removed, and the decimation procedure continues until (i) or (iii) appears.
Only when case (iii) appears, the algorithm calls WalkSAT, because the residual formula should be easy to treat. WalkSAT then tries to solve the residual formula and eventually builds the complete solution of the problem.  
Different from the other analyzed algorithms, SP is implemented in such a way that it does not output a low-energy solution to the problem when case (i) is encountered. This is a limitation of the present implementation of SP. 

Each SP iteration requires $O(N)$ operations, which yields $O(N t_{max})$, where $t_{max}$ is the maximum time allowed for finding a convergence, i.e., a big constant. In the implementation described above, the SID has a computational complexity of $O(t_{max} N^2 \log N )$, where the $N\log N$ comes from the sorting of the biases. This can be reduced to $O(Nt_{max}(\log N)^2)$ by noticing that fixing a single variable does not affect the SP messages significantly. Consequently, SP can be called every $N\delta$ decimation step by selecting a fraction of variables at each decimation step. 
In our analysis, we set $\delta = 0.00125$, which allows the algorithm to remain quadratic up to size $N=2048$. For sizes greater than $N=2048$, the fraction is still small enough to speed up the algorithm without compromising its efficiency.
The efficiency of the algorithm recalled above can be improved by introducing a backtracking strategy. We refer to \cite{marino_backtracking_2016} and references therein for a complete explanation of this strategy.

\subsubsection{Message Passing algorithms for $q$-col}
Given the $q$-col energy in eq. \eqref{eq:E_qcol},
the interaction between nearest neighbors $x_i$ and $x_j$ is associated to a Boltzmann weight $\psi_{ij}(x_i,x_j) \propto e^{-\beta \delta_{x_i x_j}}$ and the cavity equations read \cite{Pearl1982}
\begin{equation}
    \eta_{i \rightarrow j} (x_i) = \frac{1}{Z_{i \rightarrow j}} \prod_{k \in \partial i\backslash j} [1-(1-e^{-\beta})\eta_{k \rightarrow i}(x_i)]
\end{equation} 
where $Z_{i \rightarrow j}$ is a normalization term that ensures $\sum_{s_i=1}^q \eta_{i\rightarrow j}(x_i) = 1$.
\\We are interested in the problem of perfect coloring that correspond to the limit $\beta \rightarrow \infty$ of cavity equations
\begin{equation}
    \eta_{i \rightarrow j} (x_i) = \frac{1}{Z_{i \rightarrow j}} \prod_{k \in \partial i\backslash j} [1-\eta_{k \rightarrow i}(x_i)]
    \label{Eq:BP_col}
\end{equation}
Given an instance of the $q$-col problem, so a graph $G(V,E)$ and the number $q$ of available colors, we initialize random normalized cavity messages $\eta_{i\rightarrow j}(x_i)$ and we update them with eq. \eqref{Eq:BP_col}. A single update of all messages for each node of the graph is one Belief Propagation (BP) step. 
\\We repeat BP steps until convergence or a maximum time. In the BP algorithms for coloring, we considered convergence achieved if the sum of all absolute differences of all cavity messages $\eta_{i\rightarrow j}(s_i)$ between two iterations is lower than a threshold $\epsilon=10^{-7}$. The maximum number of iterations was fixed at $10^3$, so a fixed number for all $N$. At the end of the procedure, we can compute the distribution of marginals for each node of the graph
\begin{equation}
    \eta_i(x_i) = \frac{1}{Z_i} \prod_{j\in \partial i} \eta_{j \rightarrow i} (x_i)
\end{equation}
and we need a procedure to assign a color to each node. One procedure is Decimation, analogous to the decimation procedure used for the SP algorithm in the $K$-SAT problem, while in the Supplementary Information we also test the Reinforcement procedure, that results to be less efficient. 

The Decimation procedure consists in fixing the color of some variables after the convergence of BP procedure, according to the obtained marginals, producing a new coloring problem on the subgraph of non-fixed variables. For more details and variants see \cite{Krz_aka_a_2007} and \cite{DBLP:journals/corr/abs-0709-1667}.
\\In this work, we used the so-called \textit{maximum rule}: after BP procedure on the starting graph of $N_{t=0}=N$ nodes, we select the most polarized node that is the one with the biggest value of a marginal $\eta_i(x_i)$. We fix node $i$ to that color $x_i$, and we enforce $\eta_j (x_i)$ and $\eta_{j\rightarrow k}(x_i)$ to be zero from now on if $j$ is a neighbour of $i$. Then, we remove node $i$ from equations, and we repeat the BP procedure for the subgraph of $N_t-1$ nodes until $N_t=1$. 
\\It is possible that after some decimation assignments, we arrive at a situation in which for a given variable $i$ the marginals for all colors have been fixed to zero. In such cases, it means that the decimation algorithm will not find a solution to the problem with zero errors; we assign a random color to node $i$ and we continue the algorithm, in such a way to minimize the number of errors of the algorithm.
\\In this way, we need $N$ decimation assignments to fix all nodes, and each assignment requires a time proportional to $N$, meaning that decimation has a time complexity $\mathcal{O}(N^2)$. It is possible to fix a number of nodes proportional to $N$ with each decimation assignment to decrease time complexity to $\mathcal{O}(N)$, but we find that the ability of the algorithm to obtain solutions with zero error drastically decreases in this case


\subsection{GNNs} \label{sec:GNN}
Graph Neural Networks (GNNs) encompass a group of artificial neural networks that can learn a representation of graph data by locally aggregating the information present within said structures. They implement appropriate inductive biases that enforce geometric properties that are natural to a graph, compared to normal Neural Networks. A prime example of such a property that is also important for our purposes is learning graph-level functions that are invariant to the ordering of the nodes. Concretely, for two isomorphic graphs (that have the same number of nodes, edges, and connectivity), the output of the learned function should be identical.

To describe these architectures, it is useful to provide some additional details and notation regarding graph data. A graph $\mathbb{G}$ can be defined as a tuple $\mathbb{G} = (V, E)$ where $V$ is the set of nodes, with a cardinality $|V| = \mathcal{N}$, and $E \subset V \times V$ is the set of edges. It is common for graphs to have node features, i.e., attributes associated with each node $X \in \mathbb{R}^{\mathcal{N} \times d}$. The same can be true for edge features. The structure of a graph is usually represented through an adjacency matrix $A \in \mathbb{R}^{\mathcal{N} \times \mathcal{N}}$, where $A_{ij} \neq 0$ indicates that nodes $v_i$ and $v_j$ are connected with a weight $A_{ij}$ and $A_{ij} = 0$ otherwise. In the case of unweighted and undirected connections, the adjacency matrix is binary and symmetric, i.e., $A \in \{0,1\}^{\mathcal{N} \times \mathcal{N}}$. For exposition purposes, we will focus on the latter, but GNNs defined in this setting can be easily extended to deal with other connectivity structures, such as the ones describing CSP problems, which will be detailed later. A GNN layer can be compactly described in matrix form through the following equation:
\begin{equation} \label{eq:graphconv}
    X^{(k)} = \sigma(AX^{(k-1)}W^{(k)}),
\end{equation}
where $W^{(k)}$ represents the weights of the $k$-th layer, and $\sigma(\cdot)$ is an (optional) pointwise nonlinearity. This is, of course, a simplified and general representation that can be modified in different ways. Common examples include learnable adjacency matrices, where the adjacency matrix $A^{(k)}$ is layer-dependent, or modified shift operators (instead of simply multiplying by $A$). The above operation can be written in a message-passing form that provides a clear understanding of the nature of these learning algorithms:
\begin{equation}\label{eq:msg-passing}
    x_i^{(k)} = \theta^{(k)} \left( x_i^{(k-1)}, \bigoplus_{j \in S_1 (i)} \, \phi^{(k)}\left(x_i^{(k-1)}, x_j^{(k-1)},e_{ji}\right) \right),
\end{equation}
where ${x}_i^{(k)}$ denotes node features of node $x_i$ in layer $k$, $e_{ji}$ denotes (optional) edge features from node $j$ to node $i$, $\bigoplus$ denotes a differentiable permutation invariant function (e.g., sum, mean, min, or max), $S_1(i)$ denotes the 1-hop neighbors of node $i$, and $\phi, \theta$ denote differentiable and (optionally) nonlinear functions such as Multi Layer Perceptrons (MLPs).

\paragraph{GNNs as SAT solvers.} GNNs are of interest as SAT solvers because any given SAT problem formalized through a CNF formula can be translated into a bipartite, undirected graph \cite{biere2009handbook}. Due to the NP-completeness of the SAT problem, solutions to other CSP problems can be solved through these networks by reducing the problem of interest (such as $q$-col) into a SAT problem and then building the corresponding bipartite graph. A bipartite graph is one whose vertices are partitioned into two separate and non-overlapping sets $V$ and $V'$, and every edge connects a vertex in $V$ to one in $V'$. Given a SAT problem with $N$ variables and $M$ clauses, we consider a bipartite graph formulation represented through a $\{0,1\}^{2N \times M}$ incidence matrix. An edge in such a graph indicates the presence of a literal in a clause. Concurrent work in GNN benchmarking for SAT problems refers to this structure as a Literal-Clause Graph (LCG)~\cite{li_g4satbench_2023}. Following this structure, the authors of~\cite{li_g4satbench_2023} provide the following general message-passing scheme:
\begin{equation}
\label{eq:gnn_lcg}
\begin{aligned}
h_c^{(k)} &= \text{UPD}\left(\underset{l \in S_1(c)}{\text{AGG}}\left(\left\{\text{MLP}\left(h_l^{(k-1)}\right)\right\}\right), h_c^{(k-1)}\right), \\
h_l^{(k)} &= \text{UPD}\left(\underset{c \in S_1(l)}{\text{AGG}}\left(\left\{\text{MLP}\left(h_c^{(k-1)}\right)\right\}\right), h_{\neg l}^{(k-1)}, h_l^{(k-1)}\right),
\end{aligned}
\end{equation}
where $h_l$ and $h_c$ are $d$-dimensional embedding for every literal and clause node respectively, $\text{MLP}$ is the multi-layer perception, $\text{UPD}(\cdot)$ is the update function, and $\text{AGG}(\cdot)$ is the aggregation function. In the following subsections we provide the specific implementations of these equations which we use as neural solvers.

\subsubsection{NeuroSAT}
In NeuroSAT~\cite{Selsam2019}, the authors propose to operate on the LCG graph and parametrize the model using two (learnable) vectors
($\uscore{L}{init}$, $\uscore{C}{init}$), three multilayer perceptrons
($\uscore{L}{msg}$, $\uscore{C}{msg}$, $\uscore{L}{vote}$) and two
layer-norm LSTMs~\cite{ba2016layer,hochreiter1997long}
($\uscore{L}{u}$, $\uscore{C}{u}$).  At every time step $t$, a
matrix $L^{(t)} \in \mathbb{R}^{2N \times d}$ contains
the embedding for the literals and a matrix $C^{(t)} \in
\mathbb{R}^{M\times d}$ contains the embedding for
the clauses initialized by tiling $\uscore{L}{init}$ and
$\uscore{C}{init}$ respectively. The LSTM hidden states for $\uscore{L}{u}$ and $\uscore{C}{u}$ are given by
$L_h^{(t)} \in \mathbb{R}^{2N\times d}$ and $C_h^{(t)} \in
\mathbb{R}^{M\times d}$, both initialized to zero. Finally, let $\Flip$ be the operator that takes the matrix $L$ and swaps each row of $L$ with the row corresponding to the literal's negation. A single iteration of NeuroSAT consists of applying the following message passing updates:

\begin{equation}
\label{eq:neurosat}
\begin{aligned}
(C^{(t+1)}, C_h^{(t+1)}) & = \uscore{C}{u}( [ C_h^{(t)}, M^\top \uscore{L}{msg}(L^{(t)}) ] ) \\
  (L^{(t+1)}, L_h^{(t+1)}) & = \uscore{L}{u}([ L_h^{(t)}, \Flip(L^{(t)}), M \uscore{C}{msg}(C^{(t+1)}) ])
\end{aligned}
\end{equation}
After $T$ iterations, $L_*^{(T)} =
\uscore{L}{vote}(L^{(T)}) \in \mathbb{R}^{2N}$ is computed, which contains a
single scalar for each literal that can be thought of as the literal's
vote. It is easy to see the correspondences between eqs. \eqref{eq:neurosat} and \eqref{eq:gnn_lcg}.

To train the network, we use both supervised and unsupervised loss functions. In the supervised case, for a given SAT problem $\rho$ we first average over the literals to get variable-level votes $\hat{y}_{\rho}^{(T)} = L_*^{(T)} \in \mathbb{R}^N$ and then minimize the binary cross-entropy loss between the predicted assignment $\hat{y}_{\rho}^{(T)}$
and the true assignment $y_{\rho}^{(T)}$, given by a SAT solver:
\begin{equation}\label{eq:suploss}
    \mathcal{L}_{\textit{Sup}} = -\frac{1}{N} \sum_{i=1}^{N} \left[y_{\rho_i}^{(T)} \log(\hat{y}_{\rho_i}^{(T)}) + (1 - y_{\rho_i}^{(T)}) \log(1 - \hat{y}_{\rho_i}^{(T)})\right].
\end{equation}

The unsupervised loss function on the other hands is defined as in ~\cite{ozolins2022goal}:
\begin{equation}
\label{eq:unsuploss}
\begin{aligned}
    V_c(x) &= 1 - \prod_{i\in \partial c^{+}}(1-x_i)\prod_{i\in \partial c^{-}}x_i, \hfill \\
    \mathcal{L}_{\text{Unsup}}(x) &= -\log\Bigl(\prod_{c\in\rho}V_c(x)\Bigr) = - \sum_{c \in \rho} \log \left(V_c(x)\right),
\end{aligned}
\end{equation}

where $\partial c^{+}$ and $\partial c^{-}$ are the sets of variables (in positive and negative form respectively) that occur in the clause $c$ of the problem $\rho$. This loss can be though of as an energy function that reaches $0$ only when the prediction $x$ is a satisfying assignment, thus its minimization can help to construct a possible satisfying assignment without supervision. We would like to conclude this subsection by stating that the loss function in eq. \eqref{eq:unsuploss} is the same one used to train the QuerySAT model \cite{ozolins2022goal}, which will be described in the following subsection.

\subsubsection{QuerySAT}
One issue with NeuroSAT~\cite{Selsam2019} is that the model only produces one assignment, which is then used to evaluate the loss at training time. Therefore, the feedback on properly minimizing the problem's energy only comes implicitly from the negative gradient of the training loss during backpropagation. In contrast, QuerySAT~\cite{ozolins2022goal} is designed to iteratively refine its solution by producing assignments multiple times and receiving explicit feedback in the form of loss values or gradients.

Specifically, QuerySAT is a step-wise recurrent neural network-based SAT solver that operates iteratively. At each step, it generates a query by proposing variable assignments, evaluates the query using the unsupervised loss function in eq. \eqref{eq:unsuploss}, and updates its internal state based on the evaluation results. Throughout the iterative process, QuerySAT produces a target solution that integrates information from all queries performed so far. The query mechanism is defined as follows:
\begin{align*}
        (q, h) &= NN_{A}(s_{r}) \\
        e &= loss(q) \\
        (s_{r+1}, l) &= NN_{B}(h,\; e,\; \nabla_q e) \\
        \mathcal{L} &= loss(l),
\end{align*}
where \( NN_A \) is the first network layer that generates the query \( q \) and hidden state \( h \) from the current state \( s_r \). The query is evaluated using an unsupervised loss function, producing evaluation results \( e \). The second layer, \( NN_B \), processes \( h \), \( e \), and (optionally) the gradient \( \nabla_q e \) to update the state \( s_{r+1} \) and produce the output logits \( l \), which are subsequently used to compute the final loss \( \mathcal{L} \). An interesting feature of this framework is that the same loss function can be used for query evaluation and model training, but the two can also differ. In the original implementation and our benchmarking experiments, we use the unsupervised loss in eq. \eqref{eq:unsuploss} for both query evaluation and model training.

Leveraging the query mechanism along with the unsupervised loss function serves two key purposes: (a) when queried with integer variable assignments, it provides feedback on the satisfiability status of the proposed solution; (b) when queried with fractional (real-valued) assignments, it uncovers structural insights about the problem instance. This functionality allows QuerySAT to extract comprehensive information about the problem’s underlying structure, semantics, and solutions. In conclusion, \text{QuerySAT}\emph{ is a more advanced version of }NeuroSAT, with the difference between the two being the aforementioned query mechanism (when both models are trained in an unsupervised fashion). We conclude this section by stating that the original code provided by the authors \cite{ozolins2022goal} did not function on SAT-reduced $q$-coloring problems for reasons we were not able to figure out. This being the case, we had to reimplement the architecture using the same framework as NeuroSAT \cite{li_g4satbench_2023}. Additional details on our implementation can be found in sec. 7 of the Supplementary Information.

\subsubsection{rPI-GNN}\label{subsec:PI-GNN}

The canonical problem in statistical physics consists in finding the properties of a highly dimensional system with $N$ variables $\underline{x}=\{x_1, x_2, \ldots, x_N\}$ for which one knows the shape of the \textit{energy} $E(\underline{x})$, provided that $N$ is large and that the possible configurations of the variables follow the Boltzmann distribution $P(\underline{x})=\exp(-\beta E(\underline{x})) / Z$. The solution should be found in general for any value of the parameter $\beta$, which is the inverse of the physical temperature. 

Having this in mind, the most common approach in physics to combinatorial optimization problems is to define a suitable energy, or cost function, and consider the zero temperature limit of the Boltzmann distribution. This line of reasoning has given several theoretical \cite{MezardParisiTSP1986, MezardSP, FranzPRLPspin2001, zdeborova2007phase, krzakala2007gibbs} and more practical \cite{KirkpatrickSA1983, braunstein2005survey, marino_backtracking_2016} applications. More recently, this has also extended to the design of new machine learning algorithms. The Refs. \cite{schuetz2022combinatorial, Schuetz_PIGNNcol_2022} introduced physically-inspired graph neural network (PI-GNN) to solve classical combinatorial optimization problems such as maximum cut and $q$-col.

The $q$-col, already described in Section \ref{sec:Ksat_qcol}, is defined over a simple graph. Each variable $x_i$ is associated with the node of index $i$ and can take all the integer values between $0$ and $q-1$, and the usual choice for the energy function is the one in eq. \eqref{eq:E_qcol}.

The algorithm in Ref. \cite{Schuetz_PIGNNcol_2022} uses a graph neural network to assign values to the variables $x_i$ while aiming to optimize the energy $E(\underline{x})$. Similarly to the aforementioned GNN-based models, this is done by following the node feature update eq. \eqref{eq:gnn_lcg} and then using a readout layer that outputs predictions for the variables. The difference is that the energy function used for training is a soft version of eq. \eqref{eq:E_qcol}, where each variable is encoded in L1-normalized vectors $\textbf{v}_i \in [0,1]^{q}$ with continuous components $v_i^{a}$ ($a=1,\ldots, q$), defined as:
\begin{equation}
 \mathcal{L}(\underline{\textbf{v}}) = J \sum_{\langle ij \rangle} \textbf{v}_i \cdot \textbf{v}_j \label{eq:loss_qcol}
\end{equation}
To recover hard assignments for the variables, the model's output is passed to the argmax function, i.e., $x_i =  \argmax_{\, a} \, v_i^a$.

In general, the design for the GNN in Ref. \cite{Schuetz_PIGNNcol_2022} admits $\kappa$ layers, although in practice they always use $\kappa=2$. At each layer $k$ there is a learnable node embedding $\textbf{h}_i^{k}$, with $k=0,\ldots, \kappa-1$ and $i = 1, \ldots, N$. The vector $\textbf{h}_i^{0}$ is set initially at random and then also updated during the learning process. During the learning process, $\textbf{h}_i^{k}$ also changes with time, which justifies the use of a new index in the notation $\textbf{h}_i^{k} \equiv \textbf{h}_i^{k, t}$. This is the embedding for the $i$-th vertex of the graph, at the $k$-th layer in the GNN, obtained after $t$ iterations of the learning process.

The authors of Ref. \cite{Schuetz_PIGNNcol_2022} recommend the GraphSAGE architecture \cite{Hamilton_GraphSAGE_2017} for the network's layers. The outputs of the last aggregation step $\textbf{h}_i^{\kappa, t}$ are set to have dimension $q$, and the application of a softmax function $\textbf{v}_i(t) = \text{softmax} \; \textbf{h}_i^{\kappa, t}$ ensures the normalization of the vectors $\textbf{v}_i(t)$. A dropout operation follows all the other aggregation steps.

 Although the original work on the PI-GNN highlights some promising results of this model, a recent independent preprint \cite{pugacheva_QRF_PIGNN_2024} introduced a crucial modification to the design that captured our attention. They concluded that the addition of a recurrence mechanism between the outputs of the last aggregation $\textbf{h}_i^{\kappa, t}$ and the embedding in the first layer $\textbf{h}_i^{0, t+1}$ noticeably improves the performance of the algorithm. Their algorithm is named QRF-GNN (QUBO-based Graph Neural Network with a Recurrent Feature). The reader can see our own direct comparison between the original PI-GNN and the QRF-GNN in the supplementary information.
 
 Unfortunately, the authors of Ref. \cite{pugacheva_QRF_PIGNN_2024} did not provide access to their code, so we had to implement our version. We will refer to it as \textit{recurrent} PI-GNN (rPI-GNN). There should be very few differences between the rPI-GNN and the QRF-GNN, but we cannot be sure about some details since we limited ourselves to interpreting and trying to follow a pseudo-code provided in Ref. \cite{pugacheva_QRF_PIGNN_2024}. The one small modification that we are aware of is discussed in detail below. 
 
 In practice, we simply added the recurrence mechanism to the original code published for the original PI-GNN \cite{Schuetz_PIGNNcol_2022}, with GraphSAGE layers. We also optimized the computation of the energy in eq. \eqref{eq:E_qcol}, with positive impact on running times.

 To help the interested readers in the process of building their own program, we give here the pseudo-code for the network's forward pass (see the Algorithm \ref{alg:rPIGNN_pseudo}). 

 \begin{algorithm}[H]
 \begin{algorithmic}[1]
    \State{\textbf{Input:}}
    \State{$G$: the graph to color}
    \State{$\textbf{r}_i$: fixed initial features for all $i=1,\ldots, N$}
    \State{$\textbf{h}_i^{2,t}$: output of the last forward pass}
    \State{$\textbf{h}_i^{0,t+1} \leftarrow [\textbf{r}_i \; \textbf{h}_i^{2,t}]$, for all $i=1,\ldots, N$}
    \For{$i=1, \ldots, N$}
    \State{${\bf n}_{\partial_i}^{t+1} \leftarrow A_\text{MEAN}^{1}\big( \, \{\textbf{h}_j^{0,t+1}  \,, \forall j \in \partial_i \} \, \big)$}
    \State{${\bf n}_i^{t+1} \leftarrow \theta\big(\, W^{1} \cdot [\textbf{h}_i^{0,t+1} \; {\bf n}_{\partial_i}^{t+1}] \, \big)$}
    \State{${\bf u}_{\partial_i}^{t+1} \leftarrow A_\text{POOL}^{1}\big( \, \{\textbf{h}_j^{0,t+1}  \,, \forall j \in \partial_i \} \, \big)$}
    \State{${\bf u}_i^{t+1} \leftarrow \theta\big(\, W^{2} \cdot [\textbf{h}_i^{0,t+1} \; {\bf u}_{\partial_i}^{t+1}] \, \big)$}
    \EndFor
    \State{$\{ {\bf n}_i^{t+1} \} \leftarrow BN^{1}\big(\, \{ {\bf n}_i^{t+1} \} \, \big)$}
    \State{$\{ {\bf u}_i^{t+1} \} \leftarrow BN^{2}\big(\, \{ {\bf u}_i^{t+1} \} \, \big)$}
    \For{$i=1, \ldots, N$}
    \State{$\textbf{h}_i^{1,t+1} \leftarrow \theta({\bf n}_i^{t+1}+ {\bf u}_i^{t+1})$}
    \State{$\textbf{h}_i^{1,t+1} \leftarrow \text{DROPOUT}\big(\, \textbf{h}_i^{1,t+1} \, \big)$}
     \State{$\textbf{h}_{\partial_i}^{2,t+1} \leftarrow A_\text{MEAN}^{2}\big( \, \{\textbf{h}_j^{1,t+1}  \,, \forall j \in \partial_i \} \, \big)$}
     \State{$\textbf{h}_{i}^{2,t+1} \leftarrow \theta\big(\, W^{3} \cdot [\textbf{h}_i^{1,t+1} \; \textbf{h}_{\partial_i}^{2,t+1}] \, \big)$}
    \EndFor
    \State{\textbf{return} $\{ \textbf{h}_{i}^{2,t+1} \}$}
\end{algorithmic}
\caption{Forward propagation of the rPI-GNN at time $t$}
 \label{alg:rPIGNN_pseudo}
\end{algorithm}

where 
\begin{itemize}
\item $\textbf{r}_i$ is a random vector with Gaussian components and dimension $d_r$.
 \item $[\textbf{a} \textbf{b}]$ is the concatenation of the vectors $\textbf{a}$ and $\textbf{b}$
 \item $\partial_i$ is the neighborhood of the node $i$ in the graph $G$.
 \item $A^{1}_{\text{MEAN}}(\cdot)$ and $A^{2}_{\text{MEAN}}(\cdot)$ perform the \textit{mean} aggregation, and $A^{1}_{\text{POOL}}(\cdot)$ performs \textit{pool} aggregation \cite{Hamilton_GraphSAGE_2017}. Each, with its own learnable parameters, receives a set of vectors and returns one aggregated vector.
 \item $W^{1}$, $W^{2}$, and $W^{3}$ are weight matrices whose components are also learnable.
 \item $BN^{1}(\cdot)$ and $BN^{2}(\cdot)$ perform batch normalization and also introduce learnable parameters \cite{IoffeBatchNorm2015}.
 \item $\theta(\cdot)$ is a non-linear function. In our case, we used the RELU.
 \item ${\bf n}_i^{t+1}$, ${\bf u}_i^{t+1}$, ${\bf n}_{\partial_i}^{t+1}$, ${\bf u}_{\partial_i}^{t+1}$, and ${\bf h}_{\partial_i}^{2, t+1}$,  are auxiliary vectors, and the operation ${\bf n}_i^{t+1} + {\bf u}_i^{t+1}$ is an element-wise sum of their components.
 \item DROPOUT$(\cdot)$ indicates the well-known dropout operation.
\end{itemize}

In fact, the only difference between the QRF-GNN and our rPI-GNN should be in the structure of the fixed inputs $\textbf{r}_i$. They tried several configurations and recommend setting $\textbf{r}_i$ as a combination of random and pagerank inputs \cite{pugacheva_QRF_PIGNN_2024}. For simplicity, we decided to take only random inputs, guided also by the results in Ref. \cite{pugacheva_QRF_PIGNN_2024}, which suggest that this detail is not of crucial importance. 
 
The hyperparameters were chosen as in Ref. \cite{pugacheva_QRF_PIGNN_2024}: dimension of the random input $d_r=10$; dimension of the hidden layer $d_h=50$; dropout $\text{dout}=0.5$; learning rate $l_r=0.014$.
 
Notice that the training process of the rPI-GNN is done independently for each instance of the problem. In this aspect, the rPI-GNN is more similar to classical algorithms, such as simulated annealing, than to traditional neural networks. There is no train/test scheme. Instead, the algorithm runs until it finds a solution or reaches a maximum number of epochs.

To obtain the best performance we could from the rPI-GNN, we chose to scale the maximum number of epochs $\text{nep}_{\text{max}}$ with the system size $N$. Indeed, as the graphs become larger, the rPI-GNN needs more epochs to converge to the solution. The Supplementary information contains some evidence that suggests scaling the number of epochs at least linearly with the number of nodes $N$. Therefore, for all $N$, we use the scaling $\text{nep}_{\text{max}}=100N$.

\subsection{Computation of the Score}
\label{sec:Score}

In order to compute the per-algorithm Scores (S) reported in Table \ref{Tab:score}, we proceed as follows. For each instance of any problem in the test set, we run all our algorithms, with hyperparameters as reported in Table \ref{Tab:score}. Moreover, we also run longer runs of FMS for both SAT and COL, with 1000N and 6250N iterations, respectively. We classify an instance as satisfiable if any of these algorithms finds a solution (i.e. a zero energy configuration).

We call $n_{\text{sat}}$ the number of satisfiable instances for each problem.
For the in-distribution ($N\leq 256$) test dataset, $n_{\text{sat}}$ is reported, along with the total number of instances, in Table \ref{Tab:nsat}.

\begin{table}[H]
\begin{center}
\begin{tabular}{c|c|c}
 & $n_{\text{tot}}$ & $n_{\text{sat}}$ \\
\hline
3-SAT & 42000 & 28656 \\
4-SAT & 21000 & 19806 \\
3-col & 20000 & 11750 \\
5-col & 20000 & 9184 \\
\end{tabular}
\end{center}
\caption{The total number of instances and the number of satisfiable ones for each (in-distribution, $N\leq256$) test dataset.}
\label{Tab:nsat}
\end{table}

The Score for each algorithm is then computed as the number of instances for which the algorithm finds a solution, divided by $n_{\text{sat}}$. 
In other words, the Score is the fraction of satisfiable instances that the algorithm is able to solve.

\section{Data and Code availability}

Dataset generators and data download links are provided at the official public repository of the project \url{https://github.com/ArtLabBocconi/RandCSPBench}. 
Implementations of all the algorithms used in the paper, either in Python, C++ or Julia, are also provided.

\section*{Acknowledgments}


We acknowledge financial support from the following MUR PNRR projects funded by the European Union - NextGenerationEU: ThermoQT (Project No. SOE0000098);  PRIN 2022 PNRR, Project P20229PBZR, CUP B53D23028410001 and CUP J53D23016170001, Mission 4, Component C2, Investment 1.1;  Project PE0000013-FAIR; “National Centre for HPC, Big Data and Quantum Computing - HPC”, Project CN\_00000013, CUP B83C22002940006, NRP Mission 4 Component 2 Investment 1.4; computational infrastructure DARIAH.IT, PON Project code PIR01-00022, National
Research Council of Italy.
This work was conducted in the
spirit of the Slow Science Manifesto (\url{slow-science.com}), advocating for collaborative and
sustainable research. 

\bibliography{references}



\section*{Supplementary material}

| SAT | $K=3$; S ↑ | $K=4$; S ↑ | $K=3$; RE ↓ | $K=4$; RE ↓ |
|---|---|---|---|---|
| NeuroSAT | 84.48 | 47.52 | 0.0032 | 0.0009 |
| QuerySAT | 92.838 | 66.57 | **0.0028** | 0.0005 |
| FMS | **99.98** | **95.15** | **0.0028** | **0.0002** |
| SA | 98.75 | 82.61 | 0.0025 | 0.0005 |
| SP + dec. | 82.28 | 60.2 | - | - |

| col | $q=3$; S ↑ | $q=5$; S ↑ | $q=3$; RE ↓ | $q=5$; RE ↓ |
|---|---|---|---|---|
| QuerySAT | 58.86 | 7.85 | 0.0028 | 0.0060 |
| rPI-GNN | 92.49 | 73.21 | 0.0046 | 0.0020 |
| FMS | **99.96** | **98.07** | **0.0027** | **0.0007** |
| SA | 90.25 | 79.45 | 0.0031 | 0.0011 |
| BP + dec. | 78.47 | 48.17 | 0.0414 | 0.0353 |

Table 5: **Score (S) and Residual Energy (RE) reached by each algorithm for large $N$.** The score in this case is the percentage of solved instances over the sat ones in the out-of-distribution dataset ($N = 512$ and $N = 1024$, for all $\alpha$ or $c$ values). An instance is considered sat if at least one among the analyzed algorithms found a solution. The ratio of sat instances $n_{\text{sat}}/n_{\text{tot}}$ for the various problems is 10480/16800 (3-SAT), 9940/12400 (4-SAT), 6284/8000 (3-col) 7196/8000 (5-col). The running steps are $1000N$ for SA, $100N$ for FMS in the K-SAT problem, and $1000N$ for SA, $625N$ for FMS in the $q$-col problem. The number of recurrent iterations is $2N$ for NeuroSAT and QuerySAT. Best results are in bold.

the performances increase with the size of the problem). For these values of $N$ the GNNs analyzed in this work do not reach such good results because they are used in this case out of distribution. However we reported the results of classical algorithms on these sizes because we believe they could be a good benchmark for future neural solvers developed for solving these problems at large sizes, going beyond the actual state-of-the-art.

### B.4 Different scalings of solving times with problem size

In this appendix, we provide evidence that the typical running times for finding solutions to hard CSP may scale differently with the problem size $N$, depending on the parameter $\alpha$ or $c$, controlling the hardness of the problem. Evidence for this claim has already been published in Ref. [38] for the SA algorithm. Here, we strengthen such a claim by showing data for the FMS algorithm solving random 5-coloring problems and providing a more direct and clear analysis.

While the data presented in the main text has been obtained by running the solving algorithms for a maximum number of iterations, here we present data obtained by running the FMS algorithm without any time limitation. This is a crucial aspect: on the one hand, without time limitations, we are sure to measure the right time to solutions; on the other hand, we would be in trouble running such a limitless algorithm on an instance which is unsat/uncol, i.e. without any zero energy configuration (as the algorithm would never stop). We solve the latter problem by using *planted* instances [83], which are as hard as the random instances, but possess at least one solution, the one planted at the time of generating the instance. We check a posteriori that the solutions found by FMS are orthogonal to the planted one in the large $N$ limit, implying the FMS algorithm has not exploited the presence of the planted solution in the range of values of $c$ we have explored.

One more difference with the data presented in the main text is the number of instances. Here we use 10,000 instances for any chosen value of $c$ to reduce the statistical uncertainty on the measurements.

In Fig. 6, we plot the typical time, measured as the number of iterations, required by FMS to find a solution in random 5-coloring problems. One iteration is defined as $N$ attempts to update a single variable. We plot the logarithm of the typical (i.e. median) time, after having checked that it coincides with the average over instances of $\log(t)$, with $t$ being the time each run takes to find a solution. We have chosen 3 values of the connectivity $c = 12.5$, $13.0$, and $13.5$, corresponding to different hardness regimes.

For $c = 12.5$ we are in the easy phase (for the FMS algorithm, but not for other algorithms), and the time to find a solution converges in the large $N$ limit to a constant, indicated by the horizontal line in the figure. The curve interpolating the data, taking finite-size corrections into account, is $8.07 - 48.4\, N^{-0.56}$.

For $c = 13.0$ we are very close to the algorithmic threshold, and now the time to solve the problem needs to be scaled with the problem size. The best fitting curve that interpolates the data, including finite-size corrections, is $0.92 \log(N) + 3.125 - 6.41 \exp(-\sqrt{N/68.6})$. The asymptotic scaling $\log(t_{\text{typ}}) \sim 0.92 \log(N) \implies t_{\text{typ}} \sim N^{0.92}$ suggests that scaling running times linearly with $N$, as done in the present work, is optimal to reach this regime of values of $c$ (which would be unreachable without scaling running time with problem size).

Finally, for $c = 13.5$ we are in the regime where super-polynomial times are required to find a solution. The data are well fitted by a simple power law, implying a growth of typical running times

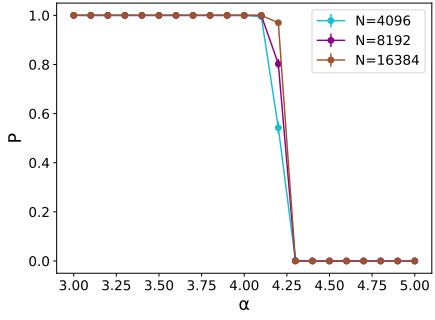

(a) SP on 3-SAT - Solving probability

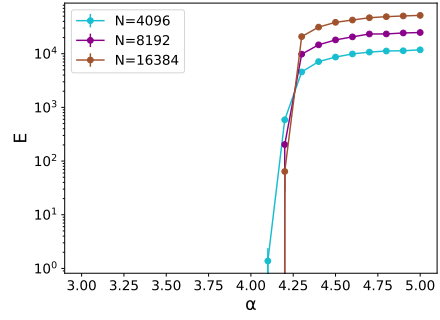

(b) SP on 3-SAT - Extensive energy

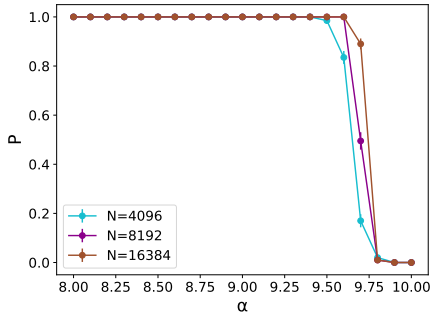

(c) SP on 4-SAT - Solving probability

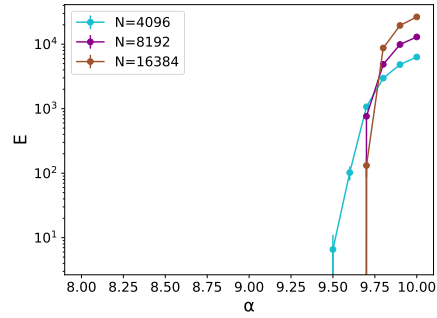

(d) SP on 4-SAT - Extensive energy

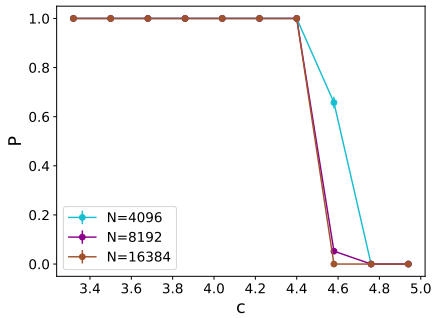

(e) FMS on 3-COL - Solving probability

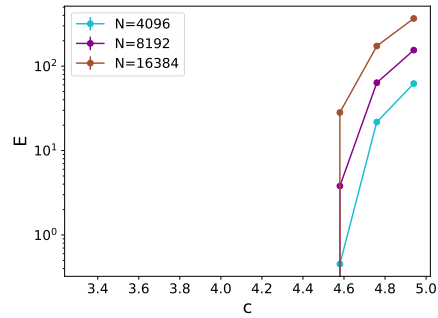

(f) FMS on 3-COL - Extensive energy

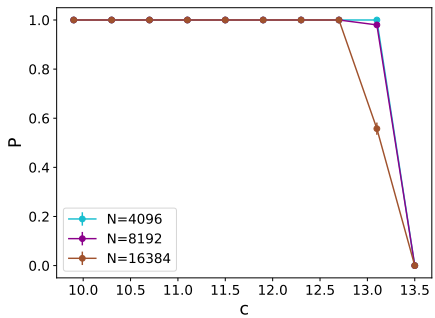

(g) FMS on 5-COL - Solving probability

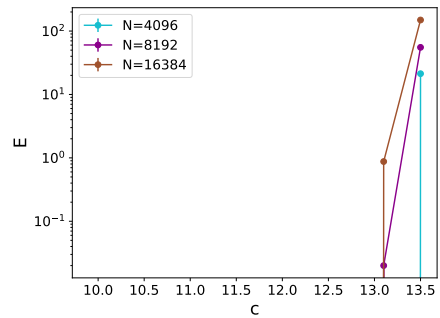

(h) FMS on 5-COL - Extensive energy

Figure 5: Solving probability and corresponding extensive energy for the FMS and SP algorithms on problems with large values of $N$.

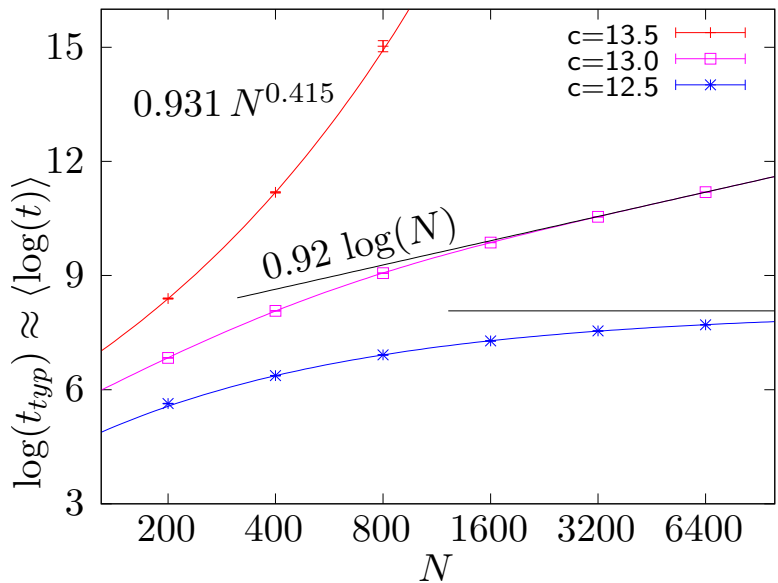

Figure 6: Typical number of iterations required by the FMS algorithm to find a solution to random 5-coloring problems at different values of the average connectivity $c$.

like $t_{\text{typ}} \sim \exp(0.931\, N^{0.415})$.

The results presented in this appendix provide direct and strong evidence that running times to find solutions in hard CSP cannot be kept constant and need to be scaled with the problem size if one wants to approach the algorithmic threshold and find solutions even in the hard regime.

### B.5 Other results for PI-GNN

This appendix includes some numerical results to support our choices for the implementation of the rPI-GNN (see subsection 3.4.3 of the main text).

First, the Fig. 7 compares the performances of the original PI-GNN [45] and the rPI-GNN in the 3-coloring problem. As in the main text, the graphs are sampled from the Erdös-Renyi ensemble and the training is done independently for each one of them. As pointed out in Ref. [81], the recurrence mechanism considerably improves the performance of the model.

On the other hand, the Fig. 8 shows that the number of epochs needed to solve the 3-coloring increases with the system size $N$. The behavior of the large systems indicates a polynomial growth with a power that depends on the connectivity. These results lead us to scale the maximum number of epochs ($\text{nep}_{\max}$), after which the rPI-GNN halts, linearly with $N$. We used the law $\text{nep}_{\max} = 100N$ for all system sizes.

### B.6 Reinforcement algorithm

Reinforcement algorithm, introduced in [84] consists in a modification of BP procedure itself, which will naturally produce greatly polarized marginals from which it is straightforward to assign colors to nodes.
We modify the Belief Propagation eq. (8) in the main text inserting reinforcement fields $\mu_i(x_i)$ to each node

$$\mu_i(x_i) = \prod_{k \in \partial i} [1 - \eta_{k \to i}(x_i)]$$
$$\eta_{i \to j}(x_i) = \frac{\mu_i(x_i)^{(1-\gamma_t)}}{Z_{i \to j}} \prod_{k \in \partial i \setminus j} [1 - \eta_{k \to i}(x_i)] \qquad (17)$$

where we have already written reinforcement fields in the limit $\beta \longrightarrow \infty$. $\gamma_t$ changes during iterations as $\gamma_t = \gamma^{t/dt}$, with $\gamma = 1 - \epsilon$ being $\epsilon$ a small, positive constant and $dt = \mathcal{O}(1)$, and $t$ the BP step we have arrived during iteration. We found $\gamma = 0.99$ and $dt = 10$ the best choice we have tried during preliminary tests. What matters is that the effect of those fields $\mu_i$ is smoothly increased during BP iterations, some different implementations of time scheduling can be found in [85] and [86].

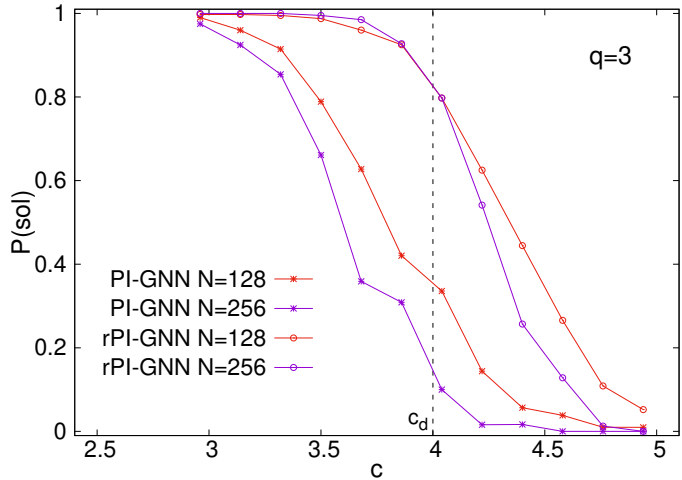

Figure 7: Comparison between the original PI-GNN and the recurrent version rPI-GNN in the 3-coloring problem. Each instance is defined over an Erdös-Rényi graph with mean connectivity $c$. The figure shows the fraction of solved instances for different mean connectivities and two system sizes: $N = 128$ and $N = 256$. The models are trained only once for each graph, with a single initial choice of the fixed random input (see Algorithm 2 in the Main text). Each point was obtained analyzing the results of between 50 (for large $c$) and 400 (for small $c$) different graphs. The dynamical spin-glass transition is known to occur at $c_d = 4$.

When $t = 0$, $(1 - \gamma_t)$ is zero and reinforcement fields have no effect. Then during BP iterations their effect increases, polarizing cavity messages up to a point where all marginals are all approximately 0 or 1, and we consequently assign colors to variables.

Since we are running BP procedure only once, fixing a maximum iteration time, $\gamma$ and $dt$, the time complexity is $\mathcal{O}(N)$ due to $\propto N$ updates for each BP step. The results of the reinforcement algorithm in this setting are reported in Fig. 9. One can see that the performances are not good compared with other analyzed algorithms. Moreover, the curves of the probability of success move towards left when $N$ is growing at variance with other good algorithms like SA, FMS or QuerySAT for which the curves cross at the algorithmic threshold for different sizes. This is probably due to the fact that the reinforcement algorithm, as introduced in this section is run for a time that scales linearly with $N$, while for any other algorithm analyzed in this paper, we let the time scaling quadratically with the system size. To use Reinforcement with a quadratic scaling it is not sufficient to run it for a maximum iteration time that scales with $N$. In fact, the behaviour of the algorithm is mainly controlled by the parameters $\gamma$ and $dt$: given chosen values for $\gamma$ and $dt$, the algorithm will produce polarized marginals after a certain average iteration time $T(\gamma, dt)$, while running it for a longer time does not improve the results. If one wants to run Reinforcement for a time that scales quadratically with the system size, one should use size dependent parameters $\gamma(N)$ and $dt(N)$ such that $T(\gamma(N), dt(N)) = \mathcal{O}(N)$. This is outside the focus of this paper, for this reason we did not include the Reinforcement algorithm in the main text.

### B.7 QuerySAT for coloring

As mentioned in the main manuscript, the original QuerySAT implementation did not work on SAT reductions of our $q$-coloring problems. The solver would output null solutions and was not able to minimize the training loss for a variety of different hyperparameters which we tested. We decided therefore to re-implement the model under the G4SATBench framework [39] which relies on PyTorch-Geometric [87] and PyTorch-Lightning [88], same as for NeuroSAT. We tried to stay as faithful as possible to the original implementation, using same feature initialization scheme, optimizer, loss normalization and gradient query mechanism. Due to the original code being written in the first version of TensorFlow, there are naturally some disparities, with the biggest one being in the gradient clipping procedure. Namely, the authors of QuerySAT use the AdaBelief optimizer [89], whose early TensorFlow implementation provides an automatic dynamic clipping procedure that can be used during training. This functionality was discontinued and not implemented in modern PyTorch versions, which are the ones we utilize, so we rely on norm-based gradient clipping, as done for NeuroSAT. Given that the gradients are also used for the queries, this can alter training

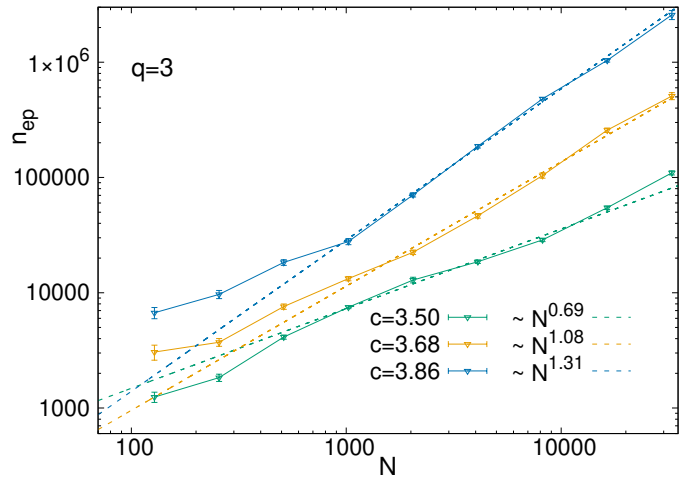

Figure 8: Average number of epochs ($n_{ep}$) needed to find a solution in the 3-coloring problem for different system sizes $N$. Both axes are in logarithmic scale. Each instance is defined over an Erdös-Rényi graph with mean connectivity $c$. The figure shows the results for three connectivities: $c = 3.5$, $3.68$, and $3.86$. The models are trained only once for each graph, with a single initial choice of the fixed random input (see Algorithm 2 of the main text). The lines are fitted only to the data points with $N \geq 1024$ and the polynomial growths have exponents $0.691 \pm 0.015$, $1.08 \pm 0.02$, and $1.31 \pm 0.02$ for the mean connectivities $c = 3.5$, $3.68$, and $3.86$, respectively. The results are averaged over 400 instances ($N \leq 8192$), or at least 10 (for $N = 32768$, $c = 3.86$). The statistics is performed only on instances that were solved by the algorithm, which is not a problem since the rPI-GNN solves almost all the studied instances when $c \leq 3.86$

dynamics, but we decided to not look into this aspect for the sake of simplicity. The complete implementation details can be found in our repository.

As shown in the main manuscript, our modern implementation is able to extract solutions on the SAT reductions of our $q$-coloring problems, thereby being more successful than the original implementation. Nevertheless, performances are quite poor, which seems to indicate that there is more room for research in the effectiveness of neural SAT solvers when dealing with reductions of other CSPs in SAT form. We believe this is a promising avenue for future work.

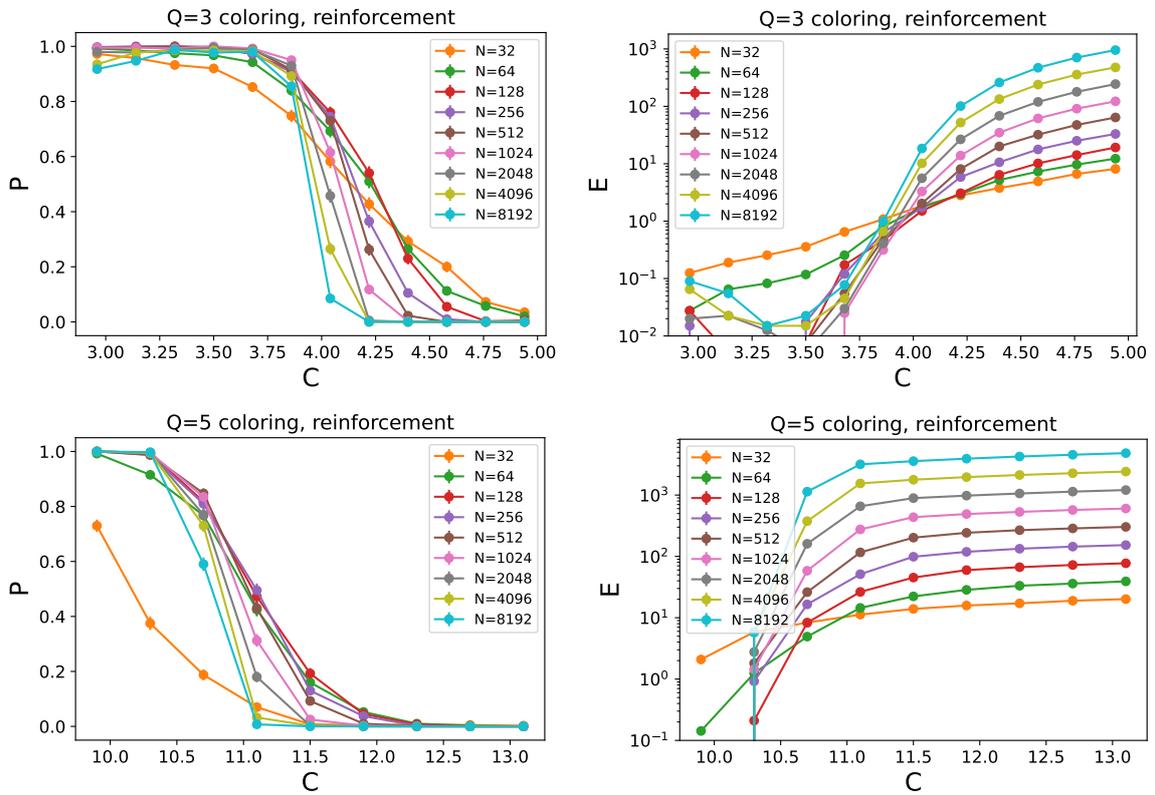

Figure 9: Results of BP + reinforcement algorithm over test dataset.



\end{document}